\documentclass[runningheads,fleqn,epsfig]{svmult}

\usepackage{makeidx}
\usepackage{graphicx}
\usepackage{multicol}
\usepackage{physmult}
\newcommand{\be}{\begin{eqnarray}}

\newcommand{\ee}{\end{eqnarray}}

\makeindex
\begin{document}
\title*{Lectures on the Theory of High Energy A+A at RHIC}
\author{Miklos Gyulassy
        \\
       {\small\it Collegium Budapest, Szentharonsag u.2, H-1014 Budapest, Hungary\\
and Physics Department, Columbia University,
New York, NY 10027} \\      
       }
\authorrunning{Miklos Gyulassy}
\maketitle
\parindent=20pt

\begin{abstract}
  In these lectures I introduce aspects of current theory used to
  interpret the preliminary data on ultra-relativistic nuclear
  collisions at RHIC energies in terms of the physical properties of
  QCD matter at extreme densities.  Topics covered include: What are
  the physics questions at SPS and RHIC?, Geometrical vs Dynamical
  features of A+A, the interplay of computable Hard pQCD vs
  phenomenological Soft dynamics, Baryon number transport and
  Junctions, How can we compute and get experimental control over the
  initial conditions?, how to reconcile apparent hydrodynamic behavior
  with partonic/hadronic transport theory.  I use the preliminary RHIC
  data available up to June 1,2001 to illustrate these topics. Most
  technical details are deferred to the literature. However, since the
  main new observable at RHIC relative to SPS is jet quenching, I
  elaborate more on this ``tomographic'' probe of ultra-dense matter.
  The possible discovery of jet quenching at RHIC by STAR and PHENIX
  is highlighted.
\end{abstract}


\section{Introduction}
Finally, after 20 years of preparation\cite{Baym:2001in}, 
a new chapter 
in nuclear/particle physics commenced on June 12, 2000 with 
the measurement of the first $Au+Au$ collisions
at $\sqrt{s}=56$ AGeV (GeV per nucleon pair)
in the Relativistic Heavy Ion Collider
(RHIC)
at the Brookhaven National Lab (BNL). 
Soon thereafter  collisions at $\sqrt{s}=130$ AGeV were also measured.
The first results were reported at Quark Matter 2001\cite{qm01} 
from the four major experiments, STAR\cite{harrisqm01}, PHENIX\cite{zajcqm01},
 PHOBOS\cite{Roland:2001me}, and BRAHMS\cite{vidabeakqm01}. 
A small army of  $\sim 1000$ experimentalists
measured the flavor, rapidity and transverse momentum
distributions of the approximately 4000 charged particles
produced in each central (head on) collision at 130 AGeV.
In  the summer of 2001, it is anticipated that RHIC
will reach its design energy, and $p+p$ and $Au+Au$
collisions at $\sqrt{s}=200$ AGeV, will come under experimental scrutiny.

These lectures provide a very condensed introduction
to current theoretical
work aimed to provide a consistent interpretation of observables
measured in such reactions in terms of the properties of dense QCD
matter. The color slides of the original lectures can be found on my
WWW site \cite{myschlad}. These lectures note are designed to
 supplement those slides and update them  with the
preliminary RHIC data available as of June 1, 2001.

The theoretical work on the new physics that may exist in QCD
matter at extreme densities
began in the mid 1970's with the realization that the asymptotic
freedom property of QCD implies the  existence of a
 new phase of strongly interacting matter
called the Quark-Gluon Plasma (QGP) \cite{collinsperry}-\cite{Braaten:1996jr}.
 Unlike familiar
nuclear or more generally hadronic matter consisting of composite
``elementary'' particles ($\pi,K,\rho,p,\Delta,\Lambda,\cdots)$ in
which quarks and gluons are permanently confined, the QGP phase
at very high temperature and/or baryon chemical potential
($T,\mu_B\gg \Lambda_{QCD}\sim 200 \;{\rm MeV}=1/{\rm fm})$
is one where the interactions between quarks and gluons become relatively
weak and short range 
$$V(r)\sim \frac{g^2}{4\pi} \frac{e^{-\mu_D r}}{r} \; ,
\;\;\alpha_s=\frac{g^2}{4\pi}\propto \frac{1}{\log(T\;{\rm or}\; \mu_B)}
\rightarrow 0\;\;.$$
The color electric (Debye) screening mass
$\mu_D(T,\mu_B)$ increases linearly with $T$ or $\mu_B$ 
modulated by a slowly varying factor of the running coupling, $g(T,\mu_B)$
(see lectures of Rebhan).
The thermodynamic properties of this deconfined and chirally
symmetric ($\sim$massless) phase of matter are thus expected in perturbation theory
to reduce approximately to an ideal Stefan-Boltzmann gas of quarks and gluons.
 For the Standard Model with 3 colors and $N_f$ flavors
of ``light'' quarks relative to $T,\mu_B$ ($SU_c(3)\otimes SU_f(N_f)$), 
the Stefan Boltzmann
constant for $\mu_B=0$ is 
$$K_{SB}=\frac{3P}{T^4}=\frac{\epsilon}{T^4}=(2_s \times8_c+\frac{7}{8}\times
 2_s\times 2_{q+\bar{q}}\times 3_c \times N_f)\frac{\pi^2}{30} (1+O(g^2))\sim 12$$
taking the helicity, color, flavor, and antiquark degrees of freedom
into account. In reality the severe infrared singularities of perturbative
QCD (pQCD) lead to large non-perturbative corrections to the ideal
gas equation of state for all temperatures and chemical potentials
accessible experimentally even beyond the future Large Hadronic Collider.
 Only numerical
lattice QCD (lQCD) methods\cite{Karsch:2001vs} (see lectures of F. Karsch) 
can provide reliable
predictions for the thermodynamic properties of the QGP phase of matter.
Effective models and resumed many-body techniques (see Alfors,
T. Rebhan, J.P. Blaizot, and E. Shuryak) are however needed 
to interpret the lQCD ``data'' and provide 
physical insight especially at finite chemical potential.
However, it is sobering to recall that 
pQCD thermodynamic expansion of the pressure
in powers of $g$ shows no sign of convergence\cite{Braaten:1996jr}
even before the Linde infrared catastrophe at $O(g^6)$,
and non-perturbative corrections to the pQCD Debye mass, $\mu_D$,
remain about a factor of $\sim 3$ up to $T\sim 200$ GeV\cite{Kajantie:1997pd}.
The full theoretical understanding of the
structure of the non-abelian plasma phase of QCD therefore remains a
fundamental open
problem in physics  because it involves 
strongly correlated, nonperturbative  and possibly
turbulent dynamical features\cite{Zheng:1997mh}. 

One of the essential and  intriguing 
aspects of the QCD many-body problem is that the {\em physical} vacuum
is an extraordinarily complex coherent many body medium.
The gluon and quark condensates lower the energy density of the
physical vacuum by and amount $B\sim \Lambda_{QCD}^4\sim 200\;{\rm
MeV/fm}^3$. Drilling a perturbative vacuum bubble of volume $V$
in this condensate costs an enormous energy $BV$. The QGP, if formed
in $V$, must counteract the physical vacuum pressure, $B$ to prevent
it from collapsing. This is only possible when the temperature exceeds
$T_c\approx (3B/K_{SB})^{1/4} \sim 150$ MeV.

The driving force behind the experimental effort at CERN and BNL over the past
20 years has been to try to create the extreme conditions necessary
to produce and diagnose this new form of matter in the laboratory.
Over the past 15 years  experiments at the AGS/BNL ($\sqrt{s}=5$ AGeV)
and the SPS/CERN ($\sqrt{s}=20$ AGeV) have searched systematically
via a very large set of observables for evidence of the QGP phase
(see lectures of C. Lourenco). In these lectures I will focus
on the most recent developments in that search that has just begun at RHIC.
I must emphasize from the onset that most of the data shown here
are of {\em PRELIMINARY} nature and could  change as better
control over the  systematic errors is achieved in the next few year.
Nevertheless, the new data are so exciting that it is worth trying
a first pass to give an overview and possible interpretation.

\section{Geometry and Dynamics in A+A}
The main obstacle in interpreting
data on  collisions of finite
nuclei (at any energy) is that the matter
created  undergoes  quantum (perhaps semi-classical)
many-body {\em
  dynamics} that may be approximated by {\em thermodynamics}
only over a limited (low $p_T$) kinematic range.
Experimentalists do not have the luxury of lattice or perturbative QCD
theorists of tapping into the infinite gedanken volume
  or reservoir with a fixed
temperature and pressure. Nuclear collisions produce dense matter in a
highly dynamical environment, and the matter produced
expands anisotropically
near the speed of light. It is far from clear whether local thermal
and chemical equilibrium concepts apply, and even so, over
what domain  of the 8 dimensional $(x^\mu,p^\mu$)
phase space can they be used.

Before a collision, the partons of the two colliding nuclei 
are locked into a coherent field configuration.
The dense virtual cloud of gluons and quarks may be described in the
colinearly factorized QCD
approximation by $A$ times the known structure functions, $f_{a/p}(x,Q)$,
 of nucleons when the resolution scale is high enough $Q > 1-2$ GeV.
However, many body initial state interactions could 
lead to strong modifications
of this naive parton picture
(see lectures of L. McLerran). The nuclear QCD fields continue to
interact after the nuclear valence quark
pancakes pass through each other. The interaction
spans a space-time hyperbola over 
a proper time $\sqrt{t^2-z^2}\sim 30\;{\rm fm/c}=10^{-22}$ sec.
Then a ``miracle'' happens! The field quanta hadronize in way
that is unfortunately not well understood.  
The dense final hadronic debris
can further interact as it expands toward the detector elements.
From CERES/SPS data\cite{ceres} there is evidence
that the in-medium mass-width (spectral function)
of vector mesons may change drastically\cite{Rapp:2000ej}.

Hard probes (jets, leptons, photons, heavy quarks) are of special interest
because they provide 
``tomographic''
 tools with which can map out this evolution 
experimentally. Hard probes are effective 
``external'' tomographic probes because
they are produced with a pQCD computable initial distribution
on a much shorter time scale, $\sim 1/m_\perp$,
then the plasma formation time, $\sim 1/3 T$. Modification of their
known initial distributions therefore
provides information on the medium through which
they propagate in analogy to conventional X-ray or positron tomography
used in medicine\cite{lovas}.
The primary advantage of RHIC over lower energy machines (AGS, SPS)
is that hard pQCD 
probes are produced at RHIC
orders 
of magnitude more abundantly over a significantly 
larger kinematic range. This greatly improves their tomographic resolution
power. 

Figure 1 shows the rapid growth of high $p_T$ $Au+Au\rightarrow
\pi^0+X$ predicted
by pQCD from SPS to RHIC and LHC.
\begin{figure}
\begin{center}
\includegraphics[width=0.65\textwidth,angle=-90]{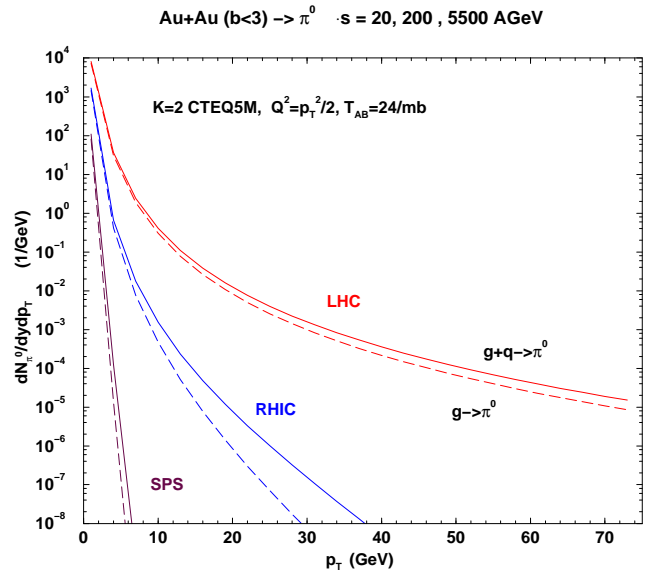}
\caption{Invariant distribution of hard pQCD produced
$\pi^0$  in central $Au+Au$
collisions as a function of c.m. 
energy via eq.(\protect{\ref{decomp},\ref{hcrossec}}) with CTEQ5M structure functions,
K=2 factor, scale $Q=p_T/2$, and multiplied by nuclear overlap $T_{AB}=24/mb$.
The dashed curve shows the contribution from gluon jet fragmentation
only. }
\label{pqcdjet}
\end{center}
\end{figure}
As discussed for example in \cite{Wang:2001bf,gvw}, it is useful to 
decompose the nuclear geometry dependence of  invariant hadron 
distributions produced in $A+B\rightarrow h+X$ at  impact parameter
${\bf b}$ into a phenomenological ``soft'' and perturbative 
QCD calculable ``hard'' components as
\begin{eqnarray}
E\frac{dN_{AB}({\bf b})  }{d^3p} &=&\; 
N_{part}({\bf b})\, \frac{dN_{soft}({\bf b}) }{dyd^2{\bf p}_{\rm T}}
+ N_{coll}({\bf b})\, \frac{1}{\sigma_{in}^{pp}}
\frac{d\sigma_{hard}({\bf b})}{dyd^2{\bf p}_{\rm T}} \; ,
\label{decomp}
\end{eqnarray}
where $N_{part}({\bf b})$ is the number of nucleon participants
and $N_{coll}({\bf b})=\sigma_{in}^{pp}T_{AB}({\bf b})$
is the number of binary $NN$ collisions at impact parameter ${\bf b}$.
The nuclear geometry of hard collisions is expressed 
in terms
of the Glauber profile density per unity area
$T_{AB}({\bf b})=\int  d^2{\bf r} \;T_A({\bf r})T_B({\bf r}-{\bf b})$
where  $T_A({\bf r})=\int dz \;\rho_A({\bf r},z)$ (see Fig.2). 
The hard part scales with the number of binary collisions $\propto A^{4/3}$
because
their probability is small built up from all possible independent
parton scattering processes. The soft part scales with only $N_{part}\propto
A^1$ because their probability is large and therefore ``shadowed''.
\begin{figure}
\vspace{-1cm}
\begin{center}
\includegraphics[width=0.75\textwidth,angle=270]{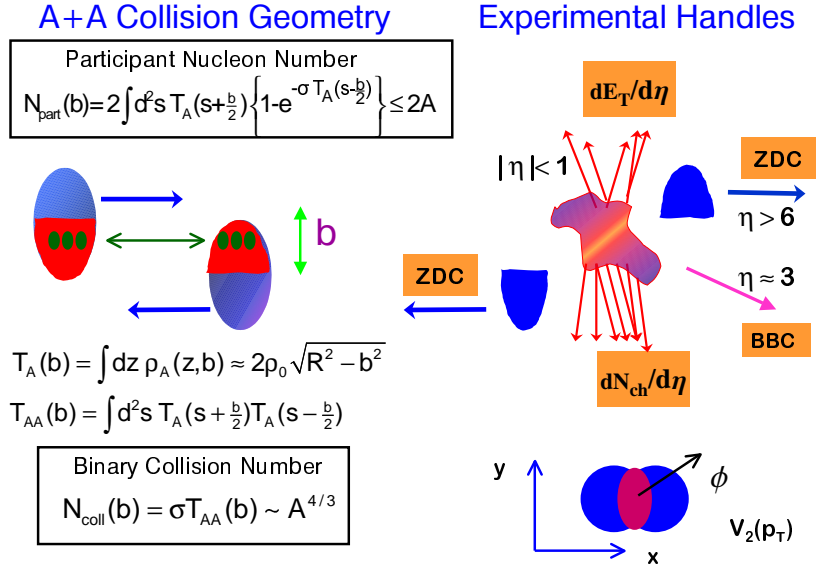}
\caption{Illustration of key aspects of the relation between
the geometry of nuclear collisions and the participant and collision number at a fixed impact parameter. The observables (see\protect\cite{harrisqm01,zajcqm01,Roland:2001me,vidabeakqm01}) used to 
constrain the geometry experimentally are also illustrated.}
\label{aageom}
\end{center}
\end{figure}

The (textbook) computable lowest order pQCD differential cross section for 
inclusive $p+p\rightarrow h+X$ invariant cross section is  given by
\begin{eqnarray}
\hspace{-4ex}
E_{h}\frac{d\sigma_{hard}^{pp\rightarrow h}}{d^3p} &=&\!
K   \sum_{abcd}\! \int\! \!dx_a 
dx_b f_{a/p}(x_a,Q^2_a) f_{b/p}(x_b,Q^2_b) 
\frac{d\sigma}{d{\hat t}}(ab\rightarrow cd)
 \frac{D_{h/c}(z_c,{Q}_c^2)}{\pi z_c}\nonumber \\ 
\; &\;&  %
\label{hcrossec}
\end{eqnarray}
where $x_a=p_a/P_A, x_b=p_b/P_B$ are the initial momentum fractions carried 
by the interacting partons, $z_c=p_h/p_c$ is the momentum fraction carried 
by the final observable hadron, $f_{\alpha/p}(x_\alpha,Q^2_\alpha)$ is the 
proton structure function for parton of flavor $\alpha$, and 
$D_{h/c}(z_c,{Q}_c^2)$ is the fragmentation function for
the parton of flavor $c$ into $h$. The UA1 data on $p\bar{p}$ hadron  
production with $p_{\rm T} > 2$~GeV can be 
well reproduced with this pQCD model expression.

The soft ($p_T<p_0\sim 2 $ GeV/c)  nonperturbative 
contribution to the hadron yields  can only
be modeled phenomenological. The Dual Parton Model
\cite{Capella:1994yb,Capella:1987cm} and the LUND string model\cite{Andersson:1983ia,Andersson:1987gw} are the most extensive and successful
low $p_T$ multiparticle phenomenologies. 
The basic pQCD matrix elements 
have been encoded into a Monte Carlo code, PYTHIA\cite{Bengtsson:1987kr}.
A variant of soft string phenomenology tuned to $pp, p\bar{p}$ data,
with the hard part taken from PYTHIA, a hadronization scheme
taken from the LUND JETSET hadronization,
and a eikonal  nuclear multiple collision geometry
were combined into the  
Monte Carlo A+B collision generator in HIJING\cite{Wang:1991ht}. 
HIJING has been used over the past decade to
predict many observables at RHIC\cite{Wang:2001bf,Wang:1991ht,Wang:1992xy}.
The separate soft and hard components in  HIJING  with a fixed $A,\sqrt{s}$
independent scale $p_0=2$ GeV/c are illustrated in Fig.3.
\begin{figure}
\vspace{-0.5cm}
\begin{center}
\includegraphics[width=0.55\textwidth,angle=270]{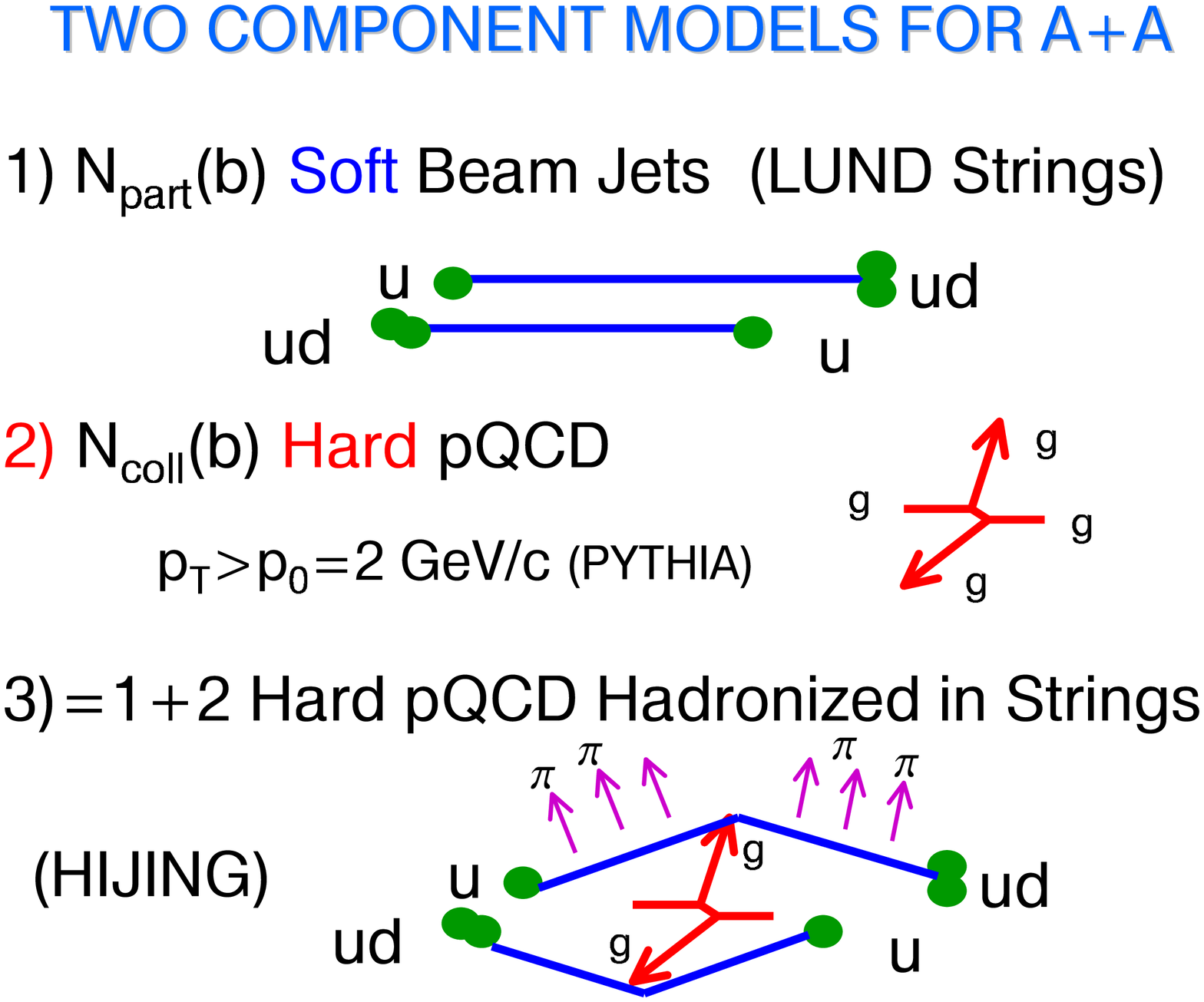}
\caption{Illustration of Hard+Soft two component models of A+A reactions
like HIJING\protect\cite{Wang:1991ht}.}
\label{twocomp}
\end{center}
\vspace{-0.75cm}
\end{figure}
Hard gluons in the LUND hadronization scheme are represented by kinks
in the strings between valence quarks and diquarks of the $N_{part}(b)$
interacting baryons in $A+B$ collisions at impact parameter b.
In this class of models no final state interactions are taken into account.

The physics in $A+A$ reactions that must be understood in order
to be able to interpret observables in terms of the
properties of dense QCD matter requires extending the above class
of event generator models to include 
\begin{enumerate}
\item Initial Conditions: 
The formation physics responsible for creating 
an incoherent gas of gluons and quarks
from the initial virtual nuclear fields,
\item Parton Transport: The $(x^\mu,p^\mu)$ phase space 
evolution of that parton gas toward equilibrium,
\item Hadronization: The dynamical mechanisms
that convert the parton degrees of freedom in
the confining physical vacuum into the observable hadronic ones,
\item Hadron Transport: The final state interactions of the expanding
dense hadronic matter
prior to ``freeze-out''.
\end{enumerate}
Each problem is fascinating in  its own right but only bits and pieces
are understood or phenomenologically mapped out up to now. 
There exists unfortunately no complete {\em computable} dynamical theory 
(like Magneto-Hydrodynamics for QED plasmas)
that consistently takes into account all four elements. QCD is believed
to be THE theory, but it is still not computable except
at high $p_T$ where perturbative or classical methods may apply.
There exists  
instead several different dynamical ``scenarios''  to describe $A+A$
that attempt to patch together different approximation techniques
and phenomenological
models to address all the physics issues in turn. 

Two generic approaches to $A+A$ can be classified by 
whether the Initial Conditions
are (1) computed (via pQCD or classical Yang Mills (cYM))
and subsequent evolution followed by a dynamical scenario for 2-4,
or (2) the initial conditions are fit by extrapolating final
observables backwards via a suitable dynamical scenario.
At lower energies (AGS,SPS) only the second approach is available
since the momentum scales are simply too low to apply either pQCD or cYM.
At collider energies RHIC and beyond, the copious production of mini-jets
\cite{Wang:2001bf,Wang:1991ht,Gaisser:1985pg,kaj,ekrt}
with $p_T>p_0\sim 2 GeV$ shown in Fig.1 
makes it possible for the first time to pursue
the first approach via pQCD Eq.(\ref{hcrossec}). At very high energies classical
Yang Mills theory\cite{mv,kmw,mkmw,mclgyu,kv,Kovchegov:2000hz} 
provides a general method to compute the Formation Physics 
which  reduces to pQCD at high $p_T$. 
Whether  RHIC or LHC energies are high enough is an open question.

The second approach, trying to ``fit'' the initial conditions
by extrapolating the final distributions backwards with a suitable dynamical
model has been traditionally based on relativistic 
hydrodynamics\cite{La53,Stocker:1986ci,Cs94,ECT97}. 
The approximate  longitudinal boost
invariant boundary conditions at ultra-relativistic energies
simplify hydrodynamic equations greatly as pointed 
out by Bjorken\cite{bjorken}. For $\mu_B=0$ the hydrodynamic equations are, 
$$\partial_\mu T^{\mu\nu}=0\;\;,\;T^{\mu\nu}(x)=u^\mu u^\nu(\epsilon+P)-g^{\mu\nu} P\;\;\;,$$
where $\epsilon(x), P(x)$ are the proper energy density and pressure
and $u^\mu(x)$ is the four velocity field of the fluid.
The central assumption is that thermal and chemical equilibrium
are maintained locally in spite of the possible large gradients
in the fluid variables. The great advantage of hydrodynamics
is that it provides a covariant dynamics depending only on the 
equation of state $P(T(x))$ that is directly related to the
lQCD predictions. When a specific space-time
freeze-out hypersurface is assumed
together with the assumption,
the Cooper-Frye 
prescription\cite{CF74,Csernai:1997xb,Bu96,ALC98,Magas:1999yb},
 the computed four fluid velocity field
can be used to predict the final  anisotropic flow pattern of hadrons. 
Since this process is assumed to be reversible, the final distributions
{\em together with an assumed freeze-out hypersurface}
can be used to compute the initial conditions on any
desired initial hypersurface. 
The disadvantage of this approach as emphasized in \cite{Molnar:2000jh}
is that
both the initial and final freeze-out hypersurfaces must be guessed.
Also  finite mean free path physics is outside the scope
of ideal hydrodynamics, and transport theory solutions\cite{Molnar:2000jh}
 do not support ``sharp'' freeze-out
hypersurfaces. Thus the inversion of data in this way
to deduce
the initial conditions is not unique. 
The neglect of dissipative effects
such a viscocity also makes it impossible to relate
central $A+A$ to peripheral and light ion data,  especially $p+p$. 
Finally, the assumption
of homogeneous or slowly spatially varying initial conditions
is questionable because of copious mini-jet production\cite{turb}.
In spite of all the above theoretical problems, initial
conditions for RHIC have been successfully  constructed 
that lead via ideal hydrodynamics and idealized Cooper-Frye freeze-out
to distributions that
reproduce amazingly well many of the low $p_T$ observables at 
RHIC\cite{rischdumi,huovinen,Kolb:2001qz,Teaney:2001gc}
(see next section). 

In order to bring the freeze-out assumption under better theoretical
control covariant, nonequilibrium transport theory\cite{degroot}
must be solved. Until recently, only simplified 1+1D Bjorken transport
theory was soluble 
in the linearized relaxation time approximation
(see \cite{Dumitru:2000up} and refs therein). This is
 due to the great numerical
complexity of the 3+1D nonlinear Boltzmann 
equations \cite{Molnar:2000jh,Zhang:1998ej}:
\newcommand{\vp}{{\bf p}}
\begin{eqnarray}
p_1^\mu \partial_\mu f_1 &=&\int\limits_2\!\!\!\!
\int\limits_3\!\!\!\!
\int\limits_4\!\!
\left(
f_3 f_4 - f_1 f_2
\right)
W_{12\to 34} \delta^4(p_1+p_2-p_3-p_4)
+  \, S(x, \vp_1) .
\label{Eq:Boltzmann_22}
\end{eqnarray}
where $W$ is the square of the $2\rightarrow 2$ scattering matrix element,
the integrals are shorthands
for $\int\limits_i \equiv \int \frac{g\ d^3 p_i}{(2\pi)^3 E_i}$,
where $g$ is the number of internal degrees of freedom,
while $f_j \equiv f(x, \vp_j)$ is the parton phase space distribution.
The initial conditions are specified by a source function $S(x,\vp)$
that corresponds to the assumed initial conditions.

Yang Pang's parton subdivision technique\cite{Zhang:1998ej,MPC}
 and the speed of current workstations
have finally made it possible to solve eq.(\ref{Eq:Boltzmann_22}) numerically.
(Codes can be obtained from the OSCAR Web site\cite{OSCAR}).
The solutions\cite{Gyulassy:1997ib,Molnar:2001ux}
 prove  that elastic parton scattering with pQCD rates
 is insufficient
at RHIC to keep the plasma in local equilibrium due to the extreme rapid
longitudinal ``Hubble'' expansion of the system\cite{Dumitru:2000up}.
Unfortunately, there exists no {\em practical} algorithm
at this time to solve the more nonlinear inelastic transport
equations involving $gg\rightarrow ng$ processes. 
Therefore, if hydrodynamics applies to $A+A$ at RHIC,
 then most likely strong 
nonperturbative mechanisms must be assumed to exist on faith
or hypothesis (see E. Shuryak lectures). This is an important
open theoretical problem.

I would also like to call attention
to a new class of hydrodynamic 
model\cite{Bass:2000ib} that side-step the final freeze-out problem
by assuming that local equilibrium is maintained
only up to an intermediate hyper-surface, just after hadronization
on a $T=T_c-\epsilon$ isotherm. Using that intermediate freeze-out as the initial conditions of a hadronic transport theory, 
the subsequent evolution of the hadronic gas toward a dynamical freeze-out
is then determined by known hadronic cross sections via
URQMD\cite{Bass:1998ca,Urqmd:1999xi}.

\section{Preliminary Results from RHIC}
\subsection{Global Constraints on Initial Conditions}
\begin{figure}
\hspace{0.0cm}
\begin{minipage}{5.5cm}
\includegraphics[width=5.5cm]{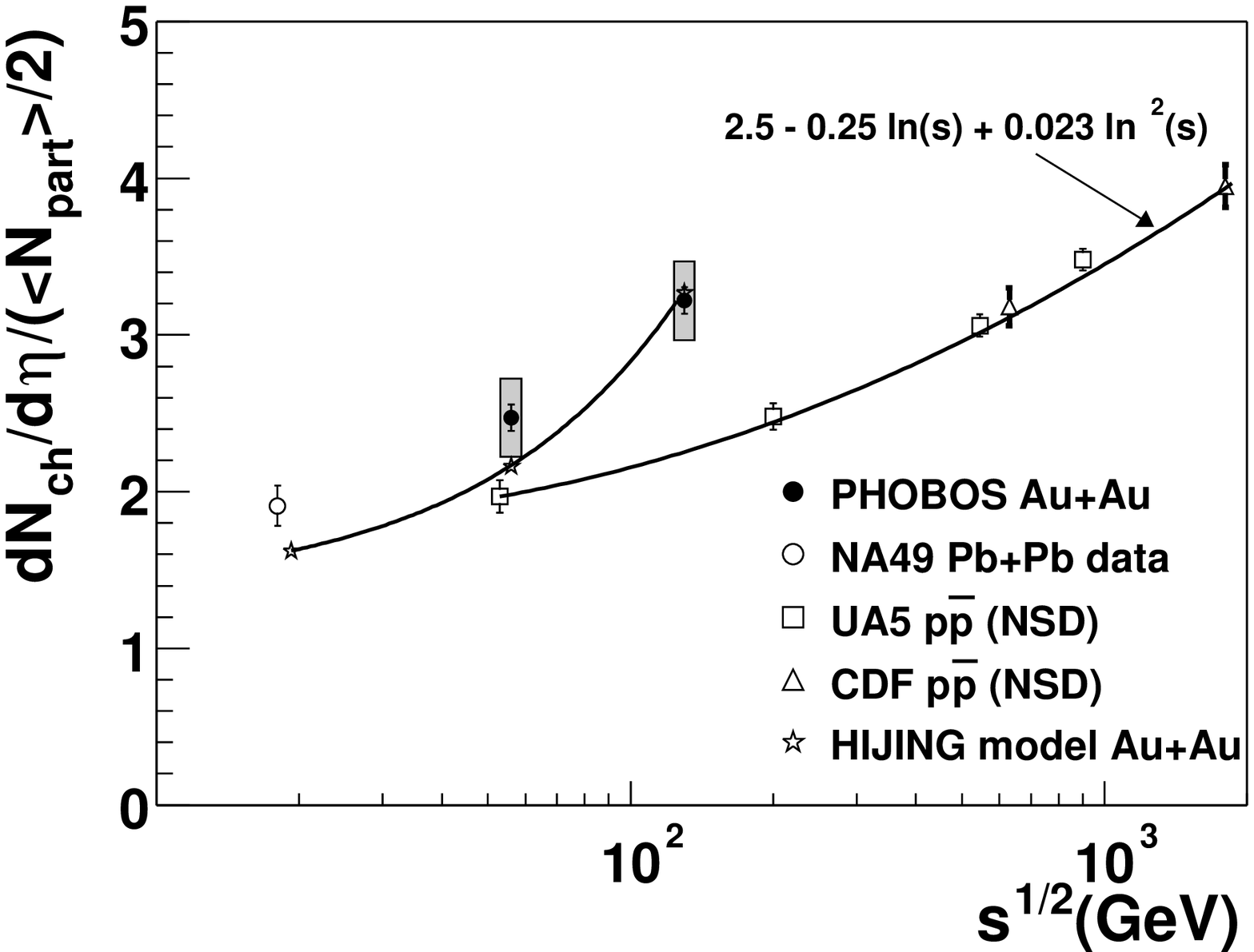}
\caption{
Measured pseudorapidity density normalized per
participant pair for central Au+Au collisions (PHOBOS\cite{Roland:2001me,Back:2000gw}).
 Systematic errors
are shown as shaded areas. 
Also shown are results of Pb+Pb data (CERN SPS), 
HIJING\protect{\cite{Wang:2001bf}} 
simulations and a parameterization of $p\overline{p}$ data.
}
\label{dndeta_vs_e}
\end{minipage}
\hspace{0.15cm}
\begin{minipage}{5.5cm}
\includegraphics[width=6.0cm,height=4cm]{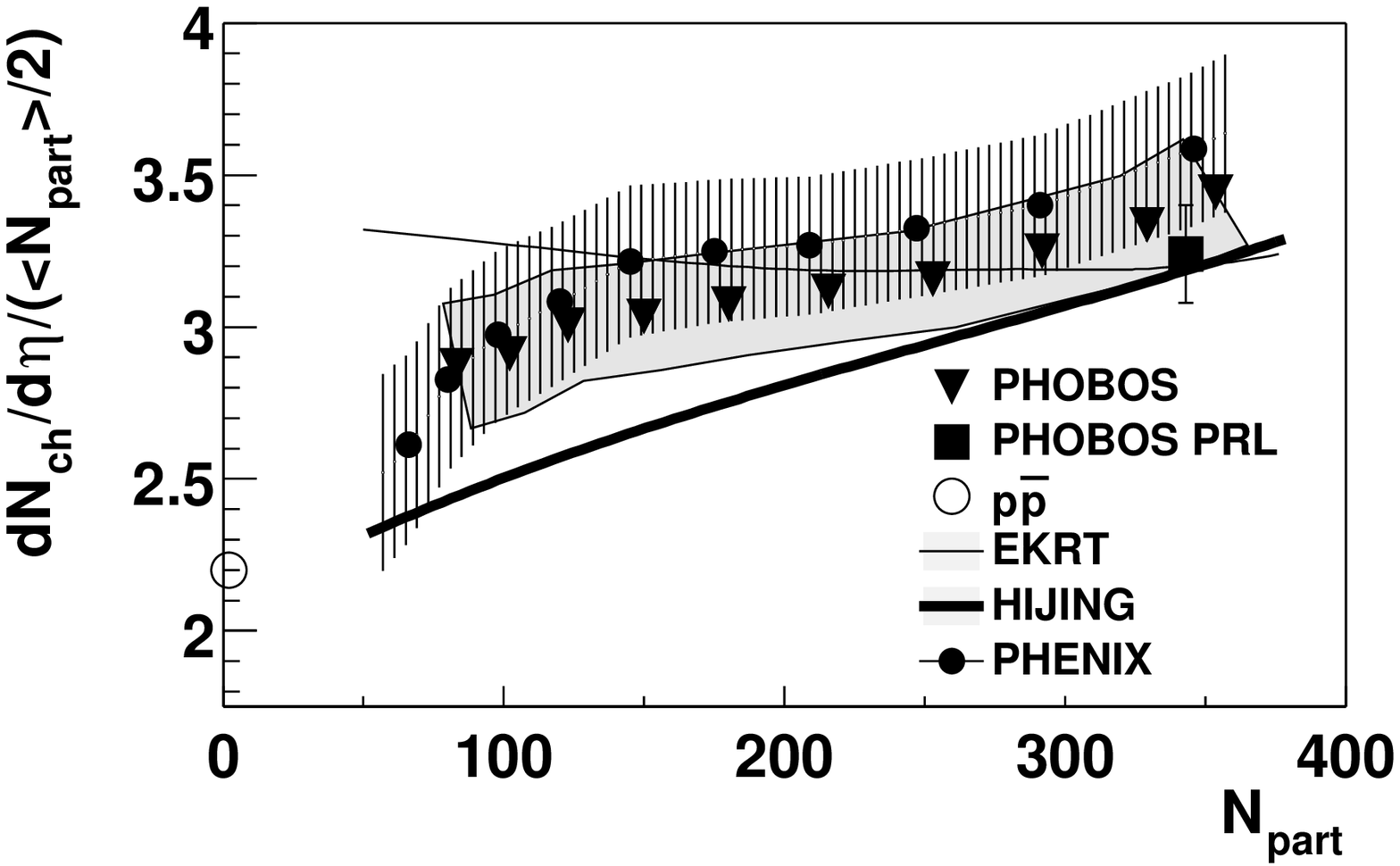}
\caption{
Normalized pseudorapidity density (PHOBOS and PHENIX)\cite{Steinberg:2001ya}
$dN_{ch}/d\eta|_{|\eta|<1}/(0.5 \times N_{part})$ as a function of the number 
of participants. Predictions based 
on HIJING (thick solid)  and EKRT\protect{\cite{ekrt}} (thin) are shown.
~~~~~~~~~~~~~~~~~~~~~~~}
\label{dndeta_vs_npart}
\end{minipage}
\vspace{-0.5cm}
\end{figure}

The first results from RHIC from PHOBOS\cite{Roland:2001me}, shown in
Fig.\ref{dndeta_vs_e}, demonstrate that the energy dependence of the
scaled charged particle (pseudo)rapidity density, $dN_{ch}/d\eta
/N_{part}$, is different from $p+p$ and $p+\bar{p}$ systematics.

Approximately 50\% more particles are produced at mid rapidity per
participating baryon in central $Au+Au$ collisions then in $p+p$ at
the same energy per baryon.  The curve shows that the two component
HIJING model predicted well this result. However, as shown in
Fig.\ref{dndeta_vs_npart}, another model EKRT\cite{ekrt}, was also 
found to predict the same multiplicity as HIJING for central
collisions. In ref.\cite{Wang:2001bf}, we proposed that the centrality
dependence of this observable could differentiate between these
competing models of the initial conditions.  The new data of PHENIX
and PHOBOS\cite{Steinberg:2001ya} verified this prediction. While
neither model accounts quantitatively
for the data,  but the two component HIJING model with its 
combined $A^1$ and $A^{4/3}$
dependence better describes the rate of increase of the scaled
multiplicity with participant number. 
The observed increase of the scaled multiplicity with energy
relative to $p+P$ and
with participant number dependence
is supports the prediction
of copious mini-jet production at RHIC.
This is one of the  necessary, though insufficient, conditions
 to form a dense gluon plasma in $A+A$.

The difference between HIJING and EKRT is that in the latter it is assume
that {\em all} the produced entropy (multiplicity)
arises at RHIC energies from hard pQCD processes.
EKRT assume that there is no significant soft component,
 i.e. $dN_{soft}\ll dN_{hard}$ in eq.(\ref{decomp}). However,
 the hard component is cutoff at scale $p_0$ that is allowed 
to vary with both $A$ and hence $N_{part}$ and 
with energy $\sqrt{s}$ based on the following assumption:
independent and hence $T_{AB}(b)$ proportional number of
gluons with $p_T>p_0$ are produced only in ``resolvable'' domains of finite
area $\pi/p_0^2$. There are $p_0^2 R^2$ such domains in the transverse
plane in a central nuclear collision.
This so called ``final state saturation'' model is then specified by
\begin{equation}
\frac{dN_g}{dy}=\frac{N_{coll}({\bf b}=0)1}{\sigma_{in}^{pp}}
\int_{p_0}^\infty d^2{\bf p}_{\rm T}
\frac{d\sigma^{A+A\rightarrow g}_{hard} }{dyd^2{\bf p}_{\rm T}}
 = \beta  p_0^2 R^2
\;\; . \label{finalsat}
\end{equation}
For $\beta=1$ assumed in EKRT, the solution for the saturation scale
is
$p_0(\sqrt{s},A)\equiv p_{sat}\approx 0.2 A^{0.13}(\surd s)^{0.19}$. 
This predicts $dN_g/dy\propto A^{0.93}$ in spite of the
apparent proportionality of hard processes to $A^{4/2}$. 
The flat $(dN_g/dy)/N_{part}\sim A^{\sim 0}$
 independence of the scaled multiplicity is a general feature of
saturating QCD models of the initial conditions (see also lectures
of Mclerran). Such a flat behavior is however ruled out by the present data
at RHIC.

An alternate (so-called initial state saturation) model was proposed by
KN\cite{dkmn} based on the nonlinear QCD evolution equations of \cite{mu}.
In this model of nuclear initial conditions, the number of liberated gluons is proportional to the number of virtual gluons participating in the
reaction on a scale $p_0$. The produced number is then taken to 
$f N_{part} xG(x,p_0)$ in terms of the nucleon gluon structure function,
where $f\sim 1.2$ is a factor on the order of unity.
Since the interaction probability is proportional to the running
coupling $\alpha_s(p_0)$, the initial state saturation condition is 
defined by
\begin{equation}
\frac{dN_g}{dy}= f N_{part}  xG(x,p_0) = 
f \frac{2}{3\pi^2\alpha_s(p_0)} p_0^2 R^2 
\; \; .\label{kharnard}
\end{equation} 
The main difference between initial and final state saturation models is
therefore due to the logarithmic dependence on 
$p_0$ introduced by the running coupling. In \cite{dkmn} a simple 
ansatz was assumed for $xG(x,Q)\propto
log Q/\Lambda$ based on the linear (DGLAP) evolution equations.
With this ansatz KN predicted a participant dependence
surprisingly close
to the observed data in Fig.\ref{dndeta_vs_npart}.

\begin{figure}
\begin{center}
\includegraphics[width=6cm,angle=-90]{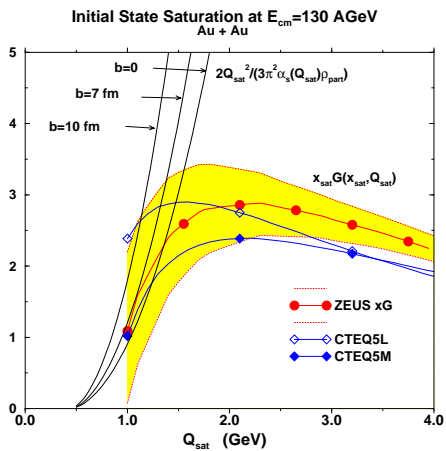}
\end{center}
\vspace{-0.5cm}
\caption{Graphical solution to the initial state
saturation eq.( \protect\ref{kharnard}) for different impact parameters
at 130 AGeV using the $xG$ gluon structure functions from ZEUS and
CTEQ. The energy dependence enters through the dependence on
$x_{sat}=2Q_{sat}/\sqrt{s}$. The participant number dependence of $Q_{sat}$
follows from the intersection of the parabolic curves with $xG$.
Unfortunately $Au$ nuclei are too small, and the solutions in the $Q\sim 1$ GeV
region are completely unreliable.} 
\label{inisat}
\vspace{-0.4cm}
\end{figure}

 However,
the $x$ independent ansatz of KN
used for $xG(x,Q)$ for the scale $Q\sim 1$ GeV/c
is a guess that cannot be supported by the 
available $ep$ HERA data. At small  $x\sim 0.01$ and low
$Q\sim 1$ the pQCD factorization analysis of deep inelastic $e+p$
 reactions
breaks down and $xG$ acquires a 100\% 
systematic error bar as shown in Fig.\ref{inisat}.
Initial state saturation is a theoretically sound model only at very high energies or nuclei with $A\gg 200$, when $Q>2$ GeV/c and the errors based on
pQCD analysis become manageable.

While it is premature to conclude which approach is least wrong
(see also \cite{wangqm01}),
in my opinion, it appears that the global multiplicity data 
and its centrality dependence can be used as indicators  that 
the initial gluon rapidity density at RHIC is between HIJING's 200 
and EKRT's 1000. The corresponding gluon density, $\rho_g(\tau)=dN_g/dy /(\tau\pi R^2) $ is thus $\sim 10-50/{\rm fm}^3$ at the corresponding formation time
$1/p_0=0.1-0.2$ fm/c. Thus RHIC may have indeed created the densest gluon
plasma ever in the laboratory. As I emphasize in a later section,
fortunately there are many other
observables, especially jet quenching,
that  provide  independent checks of this possibility.

It is important to emphasize
 that similar results for the multiplicity in central collisions 
in HIJING and EKRT are purely coincidental because the models differ
by a factor of five on the initial gluon density. 
This is compensated for by the underlying 
very different hadronization schemes assumed. 
HIJING creates a large fraction of the observed hadrons at RHIC through 
its soft string fragmentation scheme, 
while EKRT assume that entropy conservation implies that $dN_\pi\approx dN_g$.
The lack of a detailed hadronization theory can only be overcome
phenomenologically by testing experimentally all the ramifications any
particular model.

\begin{figure}
\begin{center}
\includegraphics[width=11cm,height=5.5cm]{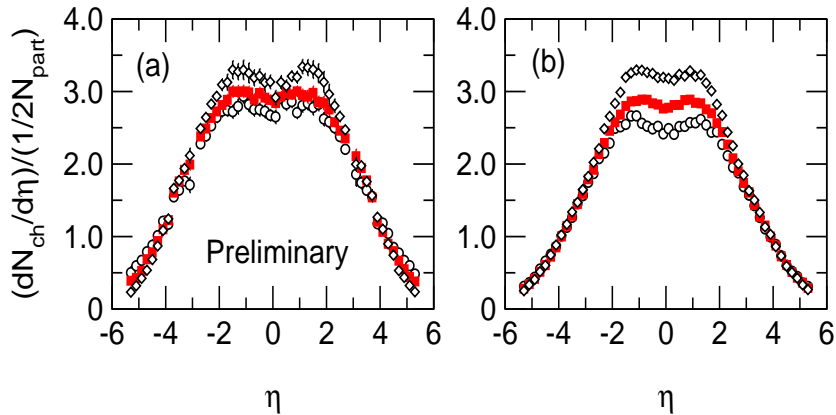}
\end{center}
\vspace{-0.5cm}
\caption{(a). Measured $dN_{ch}/d\eta/(\langle N_{part}\rangle/2)$ for $\langle N_{part} \rangle$=102 (circles), 
216 (squares) and 354 (diamonds) by PHOBOS\protect\cite{phobos_prl2}. 
(b). Same as (a) 
from HIJING.~~~~~~~~~~~~~~~~~~~~~~~~~~~~~~
}
\label{roland_phob2b}
\vspace{-0.4cm}
\end{figure}

Another observable that was suggested  in \cite{Wang:2001bf}
to help differentiate models of initial conditions
is the shape and scaling of the whole rapidity distribution 
(see Fig.\ref{roland_phob2b}).
It is seen that HIJING predicts a somewhat 
narrower  and stronger centrality dependence
than observed
by PHOBOS. This may be related to the baryon stopping power at RHIC.
Unfortunately no predictions
are available for either the initial or final saturation models
on the predicted shape of the rapidity distribution.
This observable is especially sensitive in those models to the $x$
dependence of the saturation criteria.
\begin{figure}[t]
\begin{minipage}[t]{11cm}
\begin{center}
\includegraphics[width=8cm]{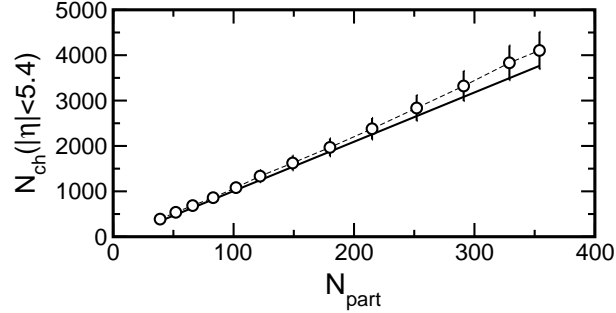}
\end{center}
\vspace*{-0cm}
\caption{\small PHOBOS total charged particle
multiplicity vs nucleon participant number\cite{Steinberg:2001ya}.}
\label{phobos_dndeta_scaled}
\end{minipage}
\hspace{\fill}
\end{figure}

The total integrate charge particle multiplicity is shown in Fig.
\ref{phobos_dndeta_scaled}. RHIC has produced about 4000 charged particles in $Au+Au$ at 130 AGeV. The nonlinear enhancement 
near central collisions is interpreted in terms of HIJING
as due to the onset of the mini-jet component.

\subsection{Global Barometric Observable \protect{$E_T/N_{ch}$}}
\begin{figure}
\begin{minipage}{12cm}
\centerline{
\includegraphics[width=8cm]{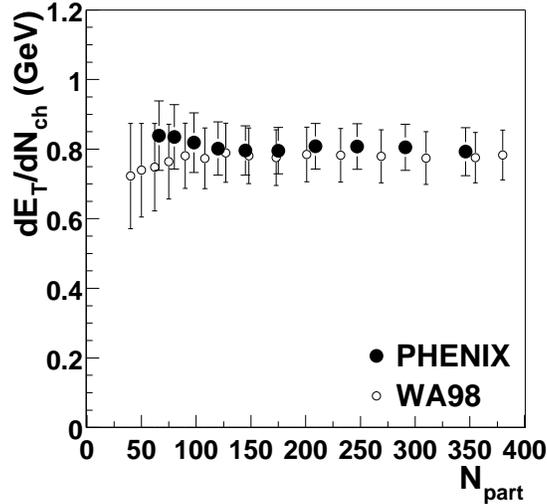}
}
\vspace*{-0.5cm}
\caption{\small Preliminary PHENIX data and WA98 data on $dE_T/dN_{ch}$ 
as a function of participant number\cite{Steinberg:2001ya}. 
This barometric observable
appears to be independent of $\sqrt{s}=20-200$
and centrality!}
\label{etnch_vs_npart}
\end{minipage}
%
%
\end{figure}
An important global barometric measure of the internal pressure
in the ultra-dense matter produced 
is the average transverse energy per charged particle.
PHENIX data are shown in Fig.\ref{etnch_vs_npart} compared to WA98 data from CERN. What is most amazing is that $E_T/N_{ch}\approx 0.8 GeV$ almost
independent of $\sqrt{s}$ from 20 to 130 AGeV and independent of centrality!
HIJING predicts that it should rise from 0.8 to 0.9 GeV from CERN to SPS
due to the enhanced mini-jet activity at RHIC. The EKRT initial state
saturation model predicts a growth of this quantity in the initial state
by about a factor of 3. The reason that EKRT remains viable after these data
is that the assumed entropy conservation implies that a large amount of $pdV$
work due to longitudinal expansion is performed by the plasma.
In 1+1D hydrodynamics the energy per particle $\epsilon/\rho \approx 2.7 T$
decrease as the system expands and cools $T\sim 1/\tau^{1/3}$.
if the freeze-out is assumed to occur at all energies and impact parameters
in $A+A$ on a fixed decoupling isotherm,
then the energy per particle will always be the same.
 At RHIC this global transverse energy loss 
from the initial state is predicted to be about a factor 3.
The theoretical problem of justifying hydrodynamics and
the freeze-out prescription itself discussed  in the previous
section comes back to haunt us here\cite{rischdumi}.
 The observed NULL effect in
$E_T/N_{ch}$ is very interesting because
it is so difficult to obtain in any transport theory with finite pQCD 
relaxation rates. 

\subsection{Discovery of Jet Quenching}

\begin{figure}
\centerline{
\includegraphics[width=2.5in,height=2.0in]{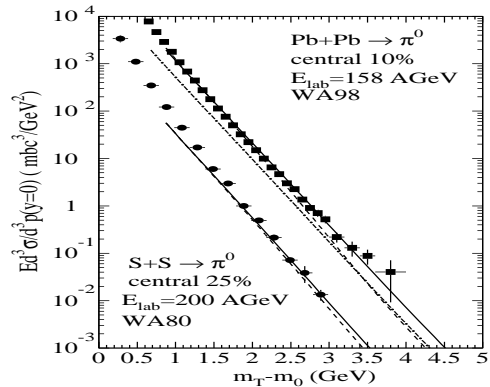}
}
\caption{ Single-inclusive $\pi^0$ spectra in central $S+S$ at 
$\protect E_{\rm lab}=200$ GeV and $Pb+Pb$ collisions at $E_{\rm lab}=158$
GeV.  The solid lines are pQCD calculations 
(Wang\protect\cite{Wang:1998hs})
with initial-$k_T$
broadening and dashed lines are without. The $S+S$ data are from WA80
and $Pb+Pb$ data are from WA98 
 The dot-dashed line is obtained from the solid line
for $Pb+Pb$ by shifting $p_T$ by 0.2 GeV/$c$.}
\label{figlet2}
\end{figure}

\begin{figure}[htb]
\begin{minipage}[t]{5.2cm}
\includegraphics[width=5.0cm, height=6cm]{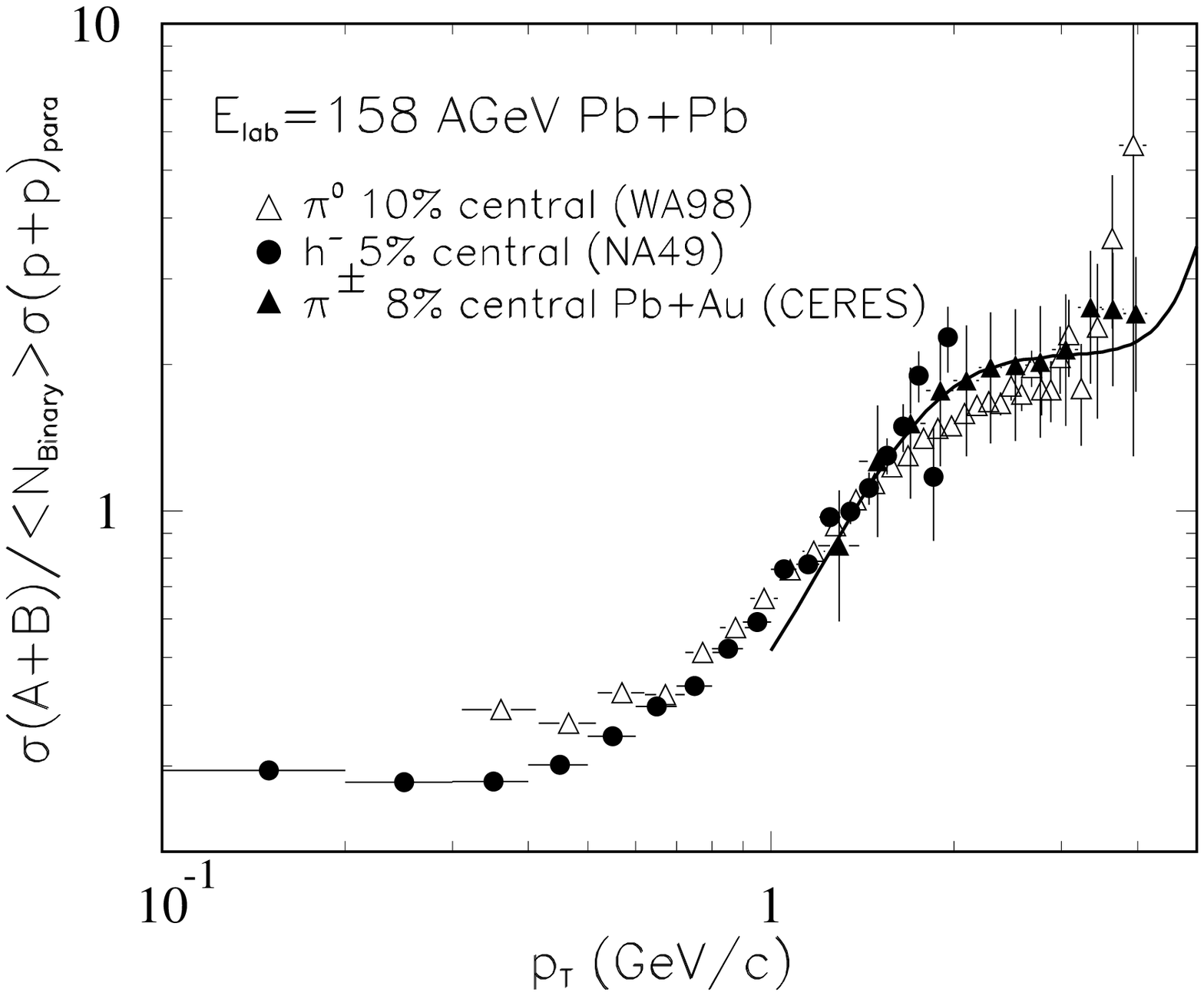}
\caption{The nuclear modification factor for hadron spectra in central
$Pb+Pb$ collisions at the CERN-SPS exceeds unity at high $p_T$
due to the Cronin effect. The solid line is a pQCD calculation 
by Wang{\protect\cite{wangqm01}}.}
\label{fig:r-sps}
\end{minipage}
\hspace{\fill}
\begin{minipage}[t]{5.8cm}
\includegraphics[width=5.5cm, height=6.5cm ]{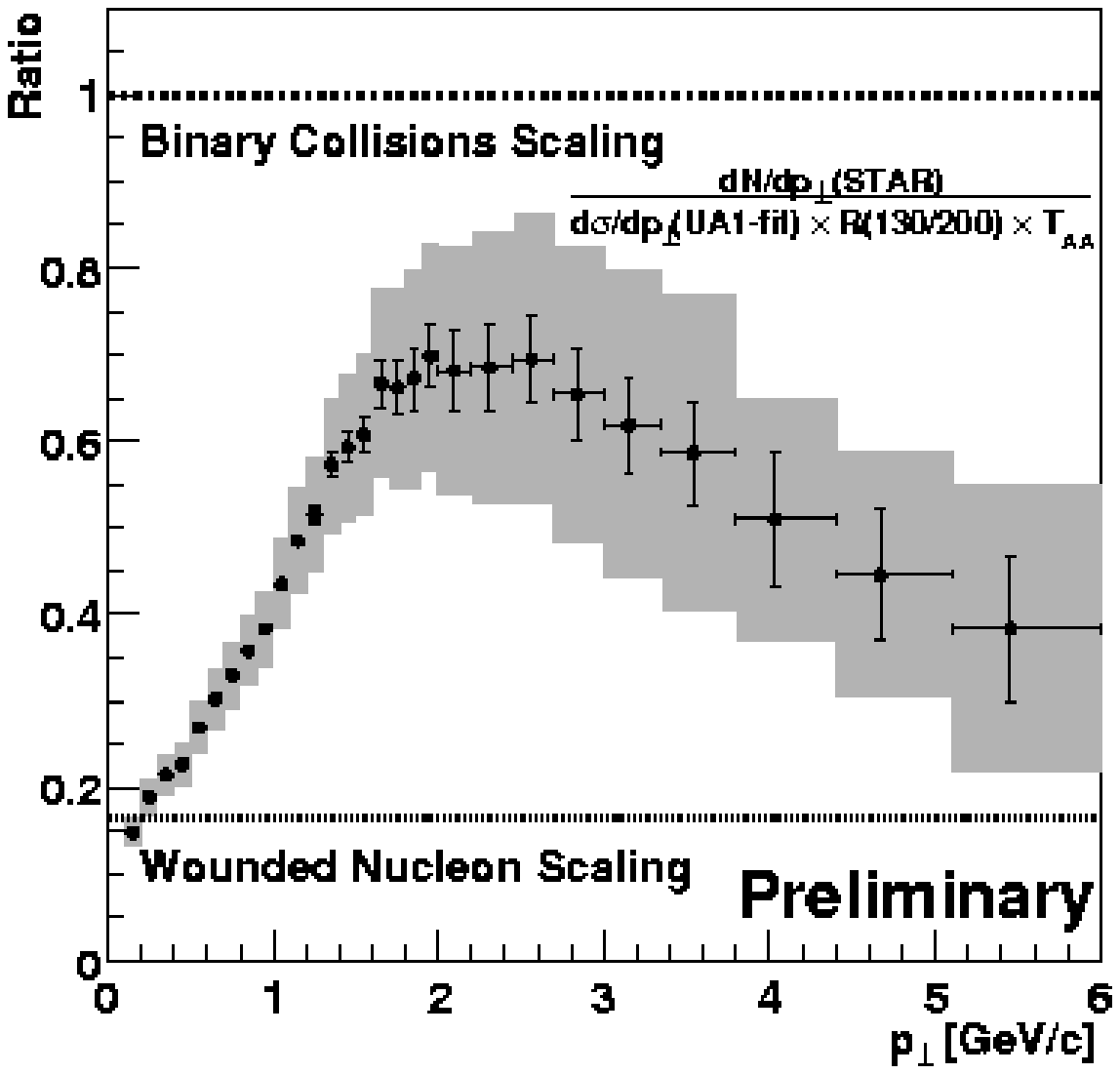}
\caption{The nuclear modification factor for charged hadrons
in central
$Au+Au$ RHIC from STAR\protect\cite{harrisqm01,dunlopqm01}.
In contrast to SPS, the high $p_T$ charged hadrons are suppressed.}
\label{fig:star_r-rhic}
\end{minipage}
\end{figure}

One of the predicted\cite{Wang:1992xy,Gyulassy:1990ye,Gyulassy:1994hr}
signatures of dense matter formation
is the suppression of jets and their high $p_T$ hadronic debris
due to energy loss of the jet in the medium.
However, the search for this effect at SPS by WA98 yielded the opposite result 
as shown in Fig.\ref{figlet2}. Even a modest $dEdx=0.2$ GeV/fm is
completely ruled out by the data\cite{Wang:1998hs}.
The problem is that at lower energies, multiple initial state elastic
scattering leads to a random walk in transverse momentum.  This
enhances the $p_T$ of the scattered partons
so that $\langle p_T^2\rangle= p_0^2 + A^{1/3} \delta p_T^2$. 
This so-called Cronin effect
has been well studied in $p+A$ reactions up to 800 GeV. At lower energies the very steep fall of the high $p_T$ tail makes the distribution 
extremely sensitive to this modest $p_T$ enhancement.
When convoluted through two nuclei, Wang predicted\cite{Wang:1998hs}
 that the Cronin enhancement
at SPS in $Pb+Pb$ should be a factor of two  as verified
in Fig.\ref{fig:r-sps}. What is plotted there is the ratio of the
observed invariant cross section to the scaled binary
collision number,  $N_{coll}(b)$, scaled invariant cross section in
$p+p$. Unity corresponds to naive superposition of $N_{coll}$ 
independent elementary
$p+p$ hard processes in the absence of any nuclear effects.
The ratio starts below 1 since the low $p_T$ distribution
grows only with the number of participants (divided by two)  and 
$N_{part}(0)/2N_{coll}(0)\approx 0.15$.

In stark contrast to the SPS enhancement of high $p_T$ pions,
 a factor 
of two or more  suppression of $p_T>2$ GeV hadrons was
reported by STAR\cite{harrisqm01,dunlopqm01} and PHENIX\cite{davidqm01}.

Fig.\ref{fig:star_r-rhic} shows that for $p_T<2$ GeV a similar trend
of increase due to the gradual transfer from participant to binary scaling
is taking place as at SPS, but for $p_T>2$ GeV the ratio for
charged particles $\pi^\pm+K^\pm+p^\pm$ starts to drop again
and reaches $\sim 0.5$ at 4 GeV/c.

\begin{figure}
\begin{flushleft}
\begin{minipage}{5.5cm}
\includegraphics[width=5.5cm,height=7cm,angle=0]{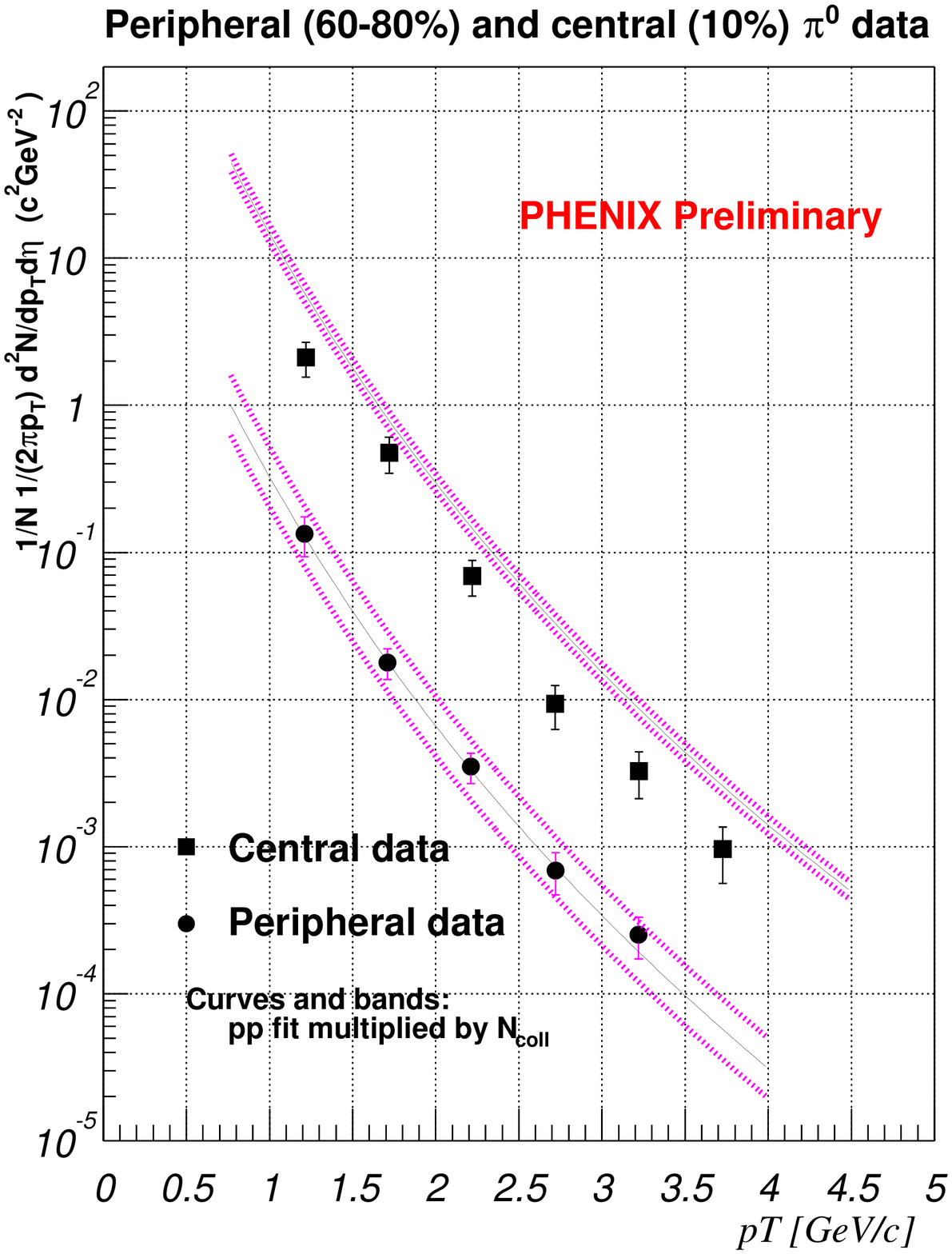}
\caption{Semi-inclusive $\pi^0$ $p_T$ distribution
        $(1/N_{int})(dN_{\pi^0}/2\pi p_Tdp_Tdy)$)
         in the upper 60-80\% peripheral events (solid
        circles) and the 10\% most central events
        (solid squares) from PHENIX\protect\cite{davidqm01}.
  The lines are a parameterization
        of $pp$ charged hadron spectra, scaled by the
        mean number of collisions $N_{coll}=857,19$ resp. 
  The bands indicate
        the possible range due to the systematic error on $N_{coll}$.
        }
\label{fig:plot_pt_ua1}
\end{minipage}
\end{flushleft}
\hspace{\fill}
\vspace{-12cm}
\begin{flushright}
\begin{minipage}[t]{5.5cm}
\includegraphics[width=5.5cm,height=7cm,angle=0]{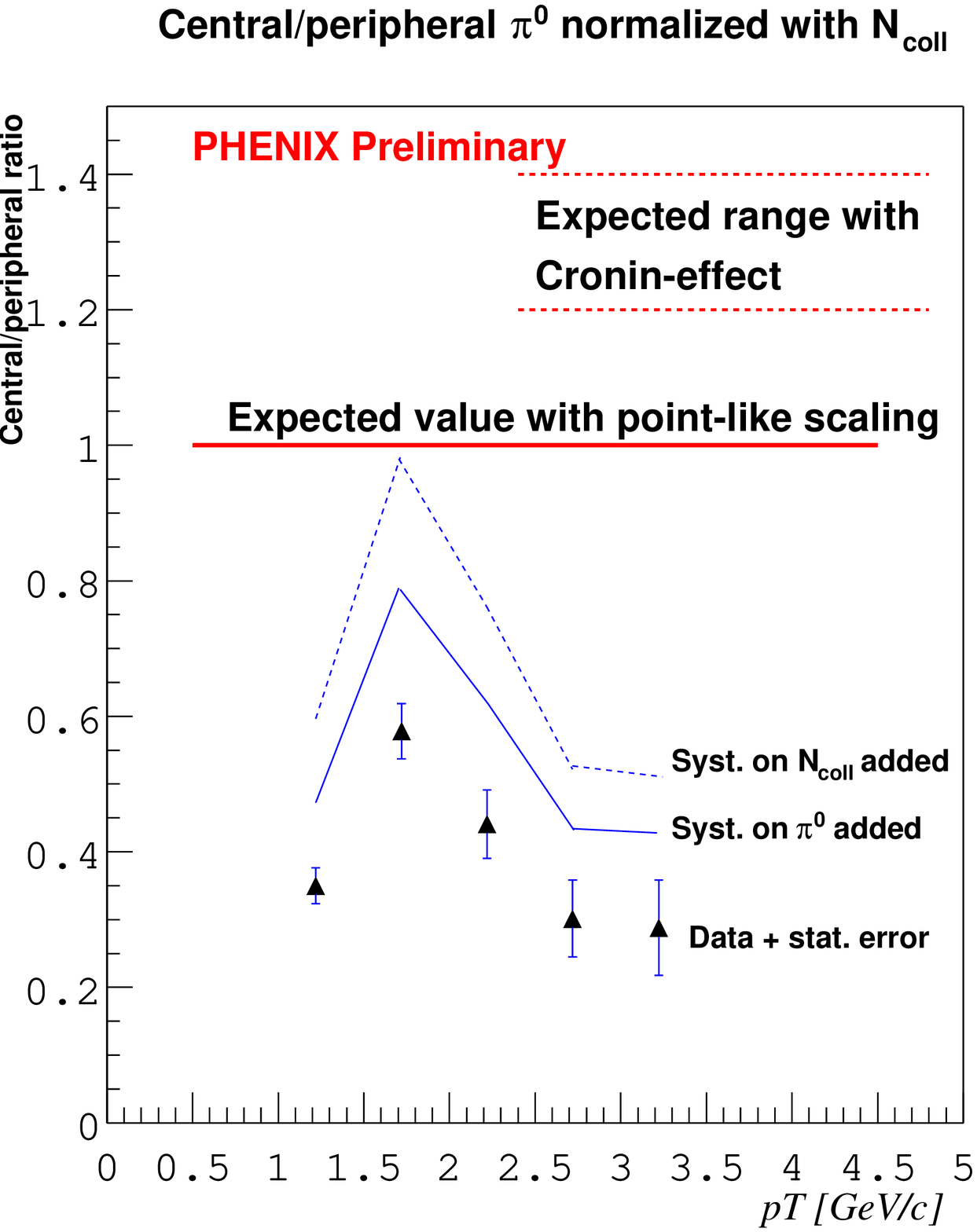}
\caption{Ratio of $\pi^0$ yields in central to peripheral
collisions at RHIC reported by PHENIX{\protect\cite{davidqm01}}
scaled by the number of binary collisions
$N_{coll}=857,19$ resp. 
The suppression of high $p_T$ pions at RHIC (in contrast to the Cronin
enhancement
at SPS in Fig.{\protect\ref{fig:r-sps}}) is due to jet quenching
in the ultra-dense matter formed at RHIC{\protect\cite{gvw,Wang:1991ht,Gyulassy:1990ye,Wang:1992xy,glv,wangqm01,levaiqm01}}
.}
\label{fig:r-rhic}
\end{minipage}
\end{flushright}
\end{figure}
The PHENIX data\cite{davidqm01} 
shows an even more dramatic quenching pattern for identified $\pi^0$
in Fig.\ref{fig:plot_pt_ua1}. In this experiment, it was further
verified that
``peripheral'' collisions are not quenched while central ones are.
Fig.\ref{fig:r-rhic} shows that the
 suppression factor may reach a factor of three at 3 GeV/c.
In this plot the ratio is not relative to pp data extrapolated to 130 GeV,
but to ``peripheral'' collisions where the average number of participants
and binary collisions is only $\approx 20$. In contrast $N_{part}\approx 360$
and $N_{coll}\approx 857$ for the central collisions. 
It must be emphasized that current systematic errors are still 
much larger than statistical, but it is clear that the combined information
from two independent experiments in 
Figs.(\ref{fig:star_r-rhic},\ref{fig:r-rhic}) imply that
something new has been discovered in $A+A$ collisions at RHIC.
I believe that this is the predicted jet quenching as discussed in
the next lecture.

The reason that this discovery is perhaps 
even more exciting than the famous
$J/\psi$ suppression effect discovered by
 NA50\cite{Abreu:2000ni,Lourenco:2001wi}
 at the SPS is that 
$J/\psi$ suppression was also seen in $p+A$. 
The cold nuclear suppression mechanism in $p+A$
is called ``normal''. The enhance suppression in $Pb+Pb$ is ``anomalous''
because it is more than  if the normal $p+A$ suppression pattern 
is extrapolated to $A+A$
That this is not theoretically fool proof was pointed out by Qiu et 
al\cite{Qiu:1998rz}. They showed that including radiative energy loss
in cold nuclei could lead to non-linear enhancement of $J/\psi$
suppression by decreasing their formation probability. 
Only a rather schematic model was presented, but it emphasizes the
necessity of improving considerably the theory of the ``normal''
processes associated with heavy quark propagation through with nuclei.
The situation is rather similar theoretically with regard to the Cronin effect.
There also only rather schematic models are available to simulate the effect.

The big difference between the two cases is that for $J/\psi$ the ``normal''
and ``anomalous'' components work in the {\em same} direction. The premium is
 thus high on developing an accurate theory  ``normal'' nuclear suppression.
In the jet quenching case, on the other hand, the
``normal'' Cronin effect works in the {\em opposite} 
direction to the ``anomalous'' new jet quenching mechanism.
Of course, there are possibly other ``normal'' effects,
 such as gluon (anti?)shadowing,
that may work in either direction at high $p_T$. 
To map out all the ``normal'' physics components
will require detailed systematic 
measurements of $p+A$ at RHIC as done at the SPS.
As a final remark, I want to emphasize 
the ``normal'' component of the dynamics is not dull run-of-the-mill
background, but fundamentally interesting
many body QCD physics in its own right and 
deserves considerable more attention.

\subsection{Where Have all the Baryons Gone?}

\begin{figure}
\centerline{
\includegraphics[width=0.65\textwidth, angle=270]{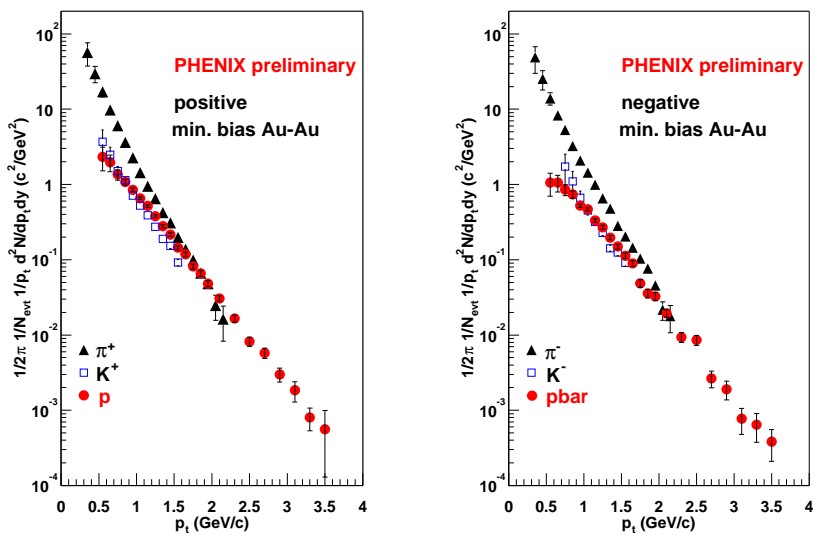}
}
\caption{
 Minimum bias transverse momentum distributions for positive
(left) and negative (right) identified hadrons measured in 
PHENIX{\protect{\cite{Velkovska:2001xz}}}. The error bars include 
statistical errors and systematic 
errors in the acceptance and decay corrections. 
Additional {\protect{$20 \%$}} systematic errors on the absolute 
normalization are not included.
}
\label{julia}
\end{figure}

One of the puzzling feature of Figs.(\ref{fig:star_r-rhic},\ref{fig:r-rhic})
is that pions appear to be more quenched than the sum of charged particles. 
Usually we assume that pions are the most abundant hadron species 
at high $p_T$ since both quark and gluon fragmentation functions 
prefer to make the lightest mesons\cite{kapiquench,Vitev:2001zn}.
Surprisingly, the preliminary PHENIX data\cite{Velkovska:2001xz}
on identified high $p_T$
hadron spectra suggest from Fig.\ref{julia} 
that baryons may be the most abundant species above $p_T>2$ GeV/c.
One possible source of such a non-pQCD like flavor distribution
could be hydrodynamic transverse flow. For a longitudinal boost invariant
(Bjorken) expansion
 with a transverse flow velocity field, $v_\perp=\tanh(\eta_r)$,
the general formula~\cite{dirk}  for the differential particle number
is 
\begin{eqnarray} 
E \frac{ dN_s }{d^3{\bf p}}  &=&  \frac{d}{2\pi^2} \int_0^1 
d \zeta \; r_f(\zeta) \tau_f(\zeta) \Bigg\{ 
-\frac{dr_f}{d\zeta}m_{\rm T} 
K_1\left(\frac{m_{\rm T} ch \eta_r}{T_f} \right)
I_0\left(\frac{p_{\rm T} sh \eta_r}{T_f} \right) 
\nonumber \\
&&+ \frac{d\tau_f}{d\zeta}p_{\rm T}
K_0\left(\frac{m_{\rm T} ch \eta_r}{T_f} \right)
I_1\left(\frac{p_{\rm T} sh \eta_r}{T_f} \right)  
\Bigg\}\;,
\label{dirf}
\end{eqnarray} 
where $d=2s+1$ is the degeneracy factor, $\eta_r={\rm Artanh}(v_\perp(z=0))$ is
the transverse fluid rapidity and $(r_f(\zeta),\tau_f(\zeta))$ is a
parameterization (counter-clockwise) of the freeze-out surface (isotherm 
of temperature $T_f$).

Solutions for freeze-out surfaces with arbitrary transverse velocity fields
$v_\perp(\xi)$ can be obtained by solving relativistic hydrodynamics. 
For the simplest case with $v_\perp =\tanh \eta_r$ a constant
and an isotherm
freeze-out on a proper time hypersurface $\tau_f$
\cite{Schnedermann:1993ws}, 
\begin{eqnarray} 
\frac{ dN_s }{dyd^2{\bf p}_{\rm T}} 
  &= & \frac{d}{4\pi^2} R^2\tau_f
m_{\rm T} 
K_1\left(\frac{m_{\rm T} \cosh \eta_r}{T_f} \right)
I_0\left(\frac{p_{\rm T} \sinh \eta_r}{T_f} \right)
\nonumber\\
&\stackrel{p_{\rm T}\gg m}{\rightarrow}& const \times d\; \exp\left(
-\frac{p_{\rm T}}{T_f \exp(\eta_r)}\right) 
\label{bjperp}
\end{eqnarray} 
which corresponds to a blue shifted effective temperature $T_f e^{\eta_r}$.
This is the uniform rapidity, transverse boosted Bjorken sausage parameterization of nuclear collision distributions.

\begin{figure}
\begin{flushleft}
\begin{minipage}{5.5cm}
\includegraphics[width=5.5cm,height=7cm,angle=0]{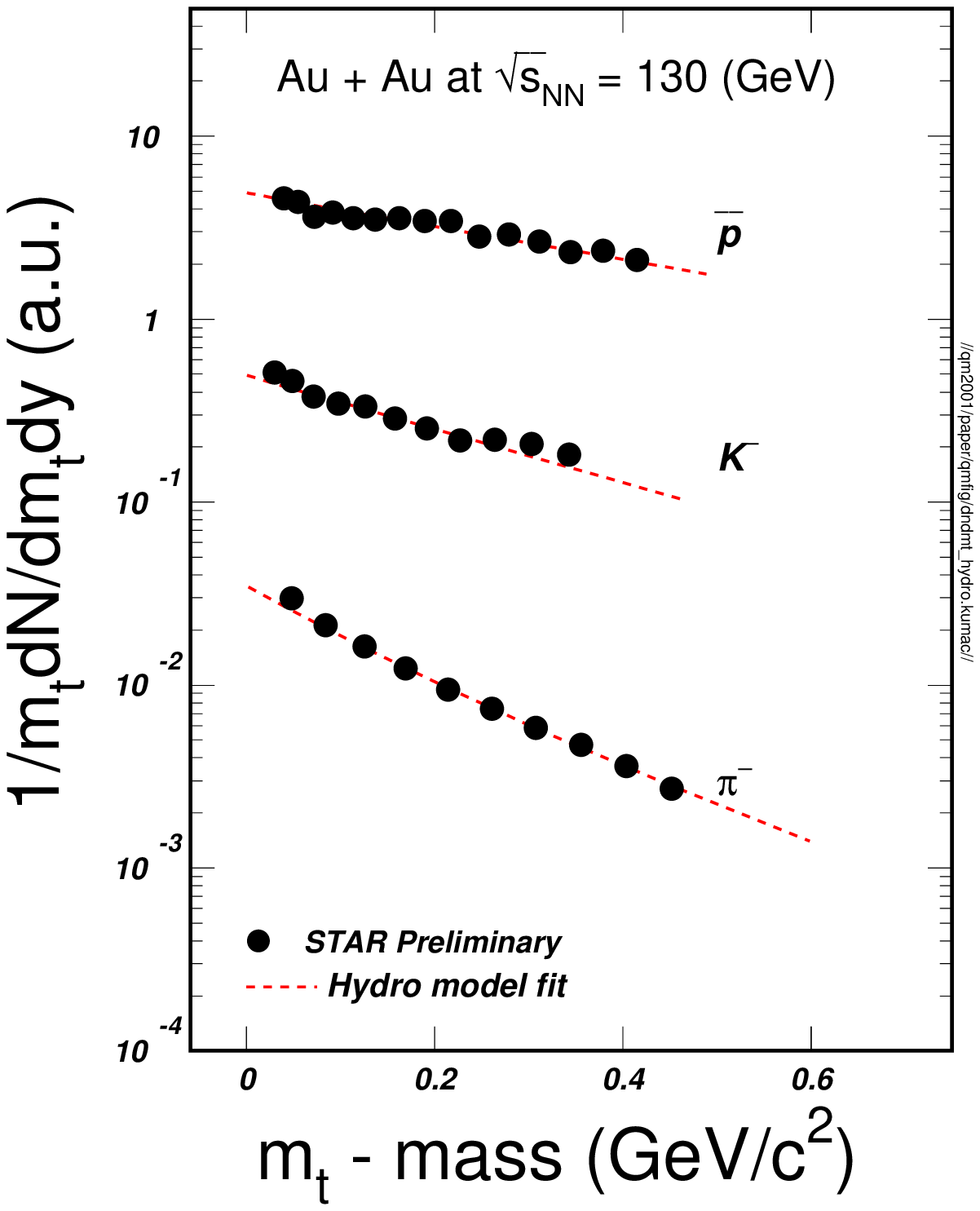}
\caption{Transverse momentum distribution for $\pi,K,p$
measured by STAR{\protect{\cite{Xu:2001zj}}} at RHIC
. The dashed curves are fits with eq.\protect\ref{bjperp}
leading to a freeze-out temperature $T_f=100$ MeV
boosted with a radial flow velocity field $v_\perp\sim 0.6$c.
}
\label{nuplot}

\end{minipage}
\end{flushleft}
\hspace{\fill}
\vspace{-10cm}
\begin{flushright}
\begin{minipage}[t]{5.5cm}
\includegraphics[width=5.5cm,height=7cm,angle=0]{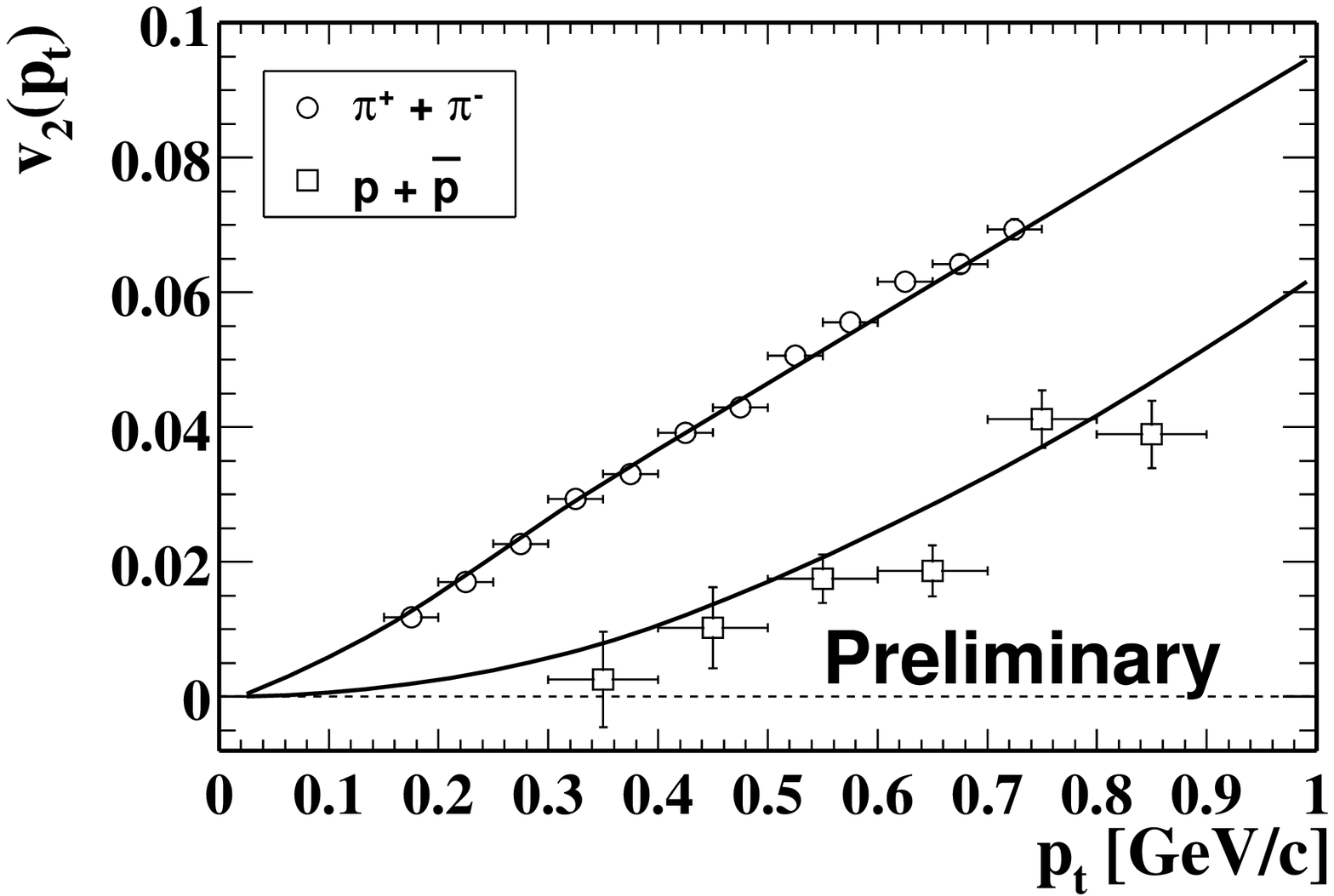}
\caption{Second azimuthal Fourier component of invariant pion and proton
distributions from STAR\protect\cite{Ackermann:2001tr} 
compared to ideal hydrodynamic
flow pattern from \protect\cite{huovinen}.}
\label{pipv2}
\end{minipage}
\end{flushright}
\end{figure}

Evidence for increased transverse flow phenomena at RHIC relative
to SPS comes from low $p_T$ STAR 
data\cite{Xu:2001zj} shown in Figs.(\ref{nuplot},\ref{pipv2}). 
The data can be fit
up to $p_T< 1$ GeV/c with a rather radial flow velocity 
$v_\perp\sim 0.6$ c
that is significantly larger than the radial flow $\sim 0.4$ c
deduced from similar SPS spectra. 

Another important experimental tool to search for collective flow effects
is to study  anisotropic multiparticle emission
patterns\cite{{Stocker:1982pg},olli92,{Reisdorf:1997fx},vpPLB}.
A particularly useful measure of collective behavior
in ultra-relativistic energies has turned out
to be the differential second Fourier component\cite{olli92} 
of the azimuthal distribution:
\begin{equation}
\frac{dN_h ({\bf b})}{dyd^2{\bf p}_{\rm T}} =
\frac{dN_h ({\bf b})}{dyd{p}^2_{\rm T}} \frac{1}{\pi}
\left(1+2 v^h_{2}(p_{\rm T})\cos(2\phi)\right)
\; ,
\end{equation}
where $\phi$ is measured relative to a ``reaction plane'' event by event
as determined  in \cite{vpPLB}. 

Azimuthal or ``elliptic'' flow results from the
initial spatial anisotropy of the dense matter in semi-peripheral
$A+A$ collisions.
The hydrodynamic model predicts an  elliptic flow pattern at RHIC
\cite{huovinen,Kolb:2001qz} that can be approximately parameterized as
\begin{equation}
v^s_{2}(p_{\rm T}) \approx {\rm tanh}(p_{\rm T}/(10\pm 2\;{\rm GeV}))
\;\; .\end{equation}
Up to about $p_T<1$ GeV, this agrees remarkably well with STAR data.
At high $p_T$ this hydrodynamic flow component
breaks down because of the emergence 
of the  hard pQCD hadrons. 

The transverse  boosted Bjorken sausage model
eq.(\ref{bjperp}) 
predicts that asymptotically the baryon/meson
ratios $p/\pi+=\bar{p}/\pi^-\rightarrow 2$
for any flow velocity because nucleons have 2 spin states.
However, numerically this ratios exceed unity only
$p_T > 3-4$ GeV.  Thus transverse flow alone cannot account for the
anomalous baryon dominance of high $p_T$ spectra in Fig.\ref{julia}
as emphasized in \cite{Vitev:2001zn}.

\begin{figure}
\begin{flushleft}
\begin{minipage}{5.5cm}
\includegraphics[width=5.5cm,height=7cm,angle=0]{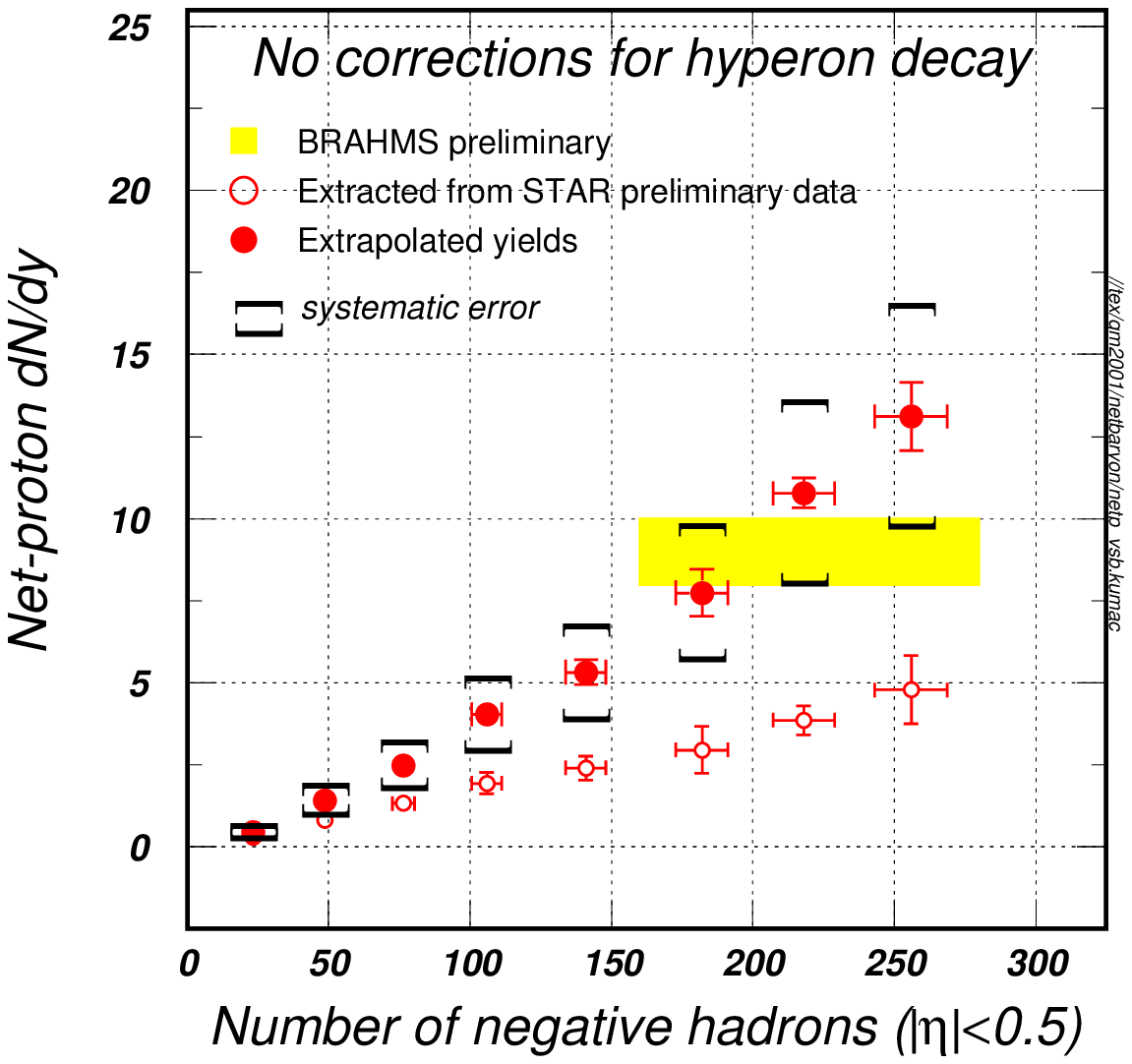}
\vspace{-1cm}
\caption{The valence proton rapidity density
measured by STAR{\protect{\cite{Xu:2001zj}}} at RHIC
as a function of $N_{part}$. 
preliminary BRAHMS data are also indicated.
}
\label{nuplot2}

\end{minipage}
\end{flushleft}
\hspace{\fill}
\vspace{-8cm}
\begin{flushright}
\begin{minipage}[t]{5.5cm}
\includegraphics[width=5.5cm,height=5cm,angle=0]{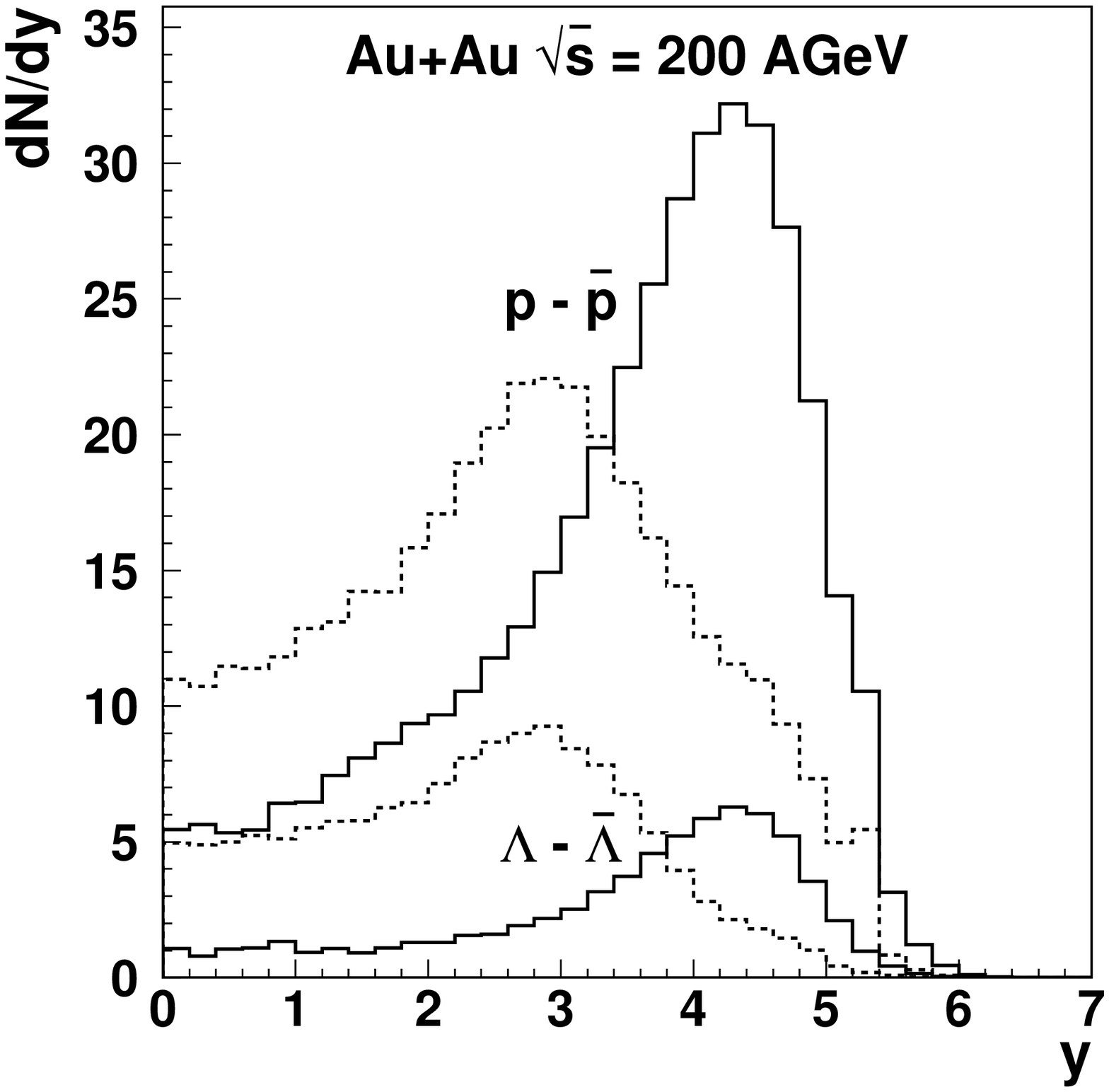}
\caption{Predicted valence proton rapidity density
at RHIC from \protect\cite{Vance:1998vh} showing factor of two
enhancement expected if baryon junction exchange is
included in HIJING/B. The dashed curves are result of HIJING 
including only standard LUND diquark fragmentation. 
}
\label{hjbbar}
\end{minipage}
\end{flushright}
\end{figure}

Another observation\cite{Xu:2001zj} that possibly provides a hint
that the answer to the puzzling result
may lie in novel baryon dynamics  at RHIC 
can be seen in Figs.(\ref{nuplot2},\ref{hjbbar}).
As was shown by Kharzeev\cite{khar_bj96}, the energy and rapidity dependence 
of the inclusive baryon production at mid-rapidity can be obtained 
using Mueller's generalized optical theorem in the 
double Regge limit.  Here, the exchanges of a Pomeron and 
a $M^J_0$ Baryon-anti-baryon ``junction''
pair lead to the following form for single mid-rapidity baryon
production,
\begin{equation}
E_B \frac{d^3\sigma^{(1)}}{d^3 p_B} = C_B f_B(m^2_t) \left ( \frac{s_0}{s}
 \right )^{1/4} \cosh(y/2)\;\; .
\label{bj1} \end{equation}
where $C_B$ is a constant that reflects the couplings
of the Reggeon and Pomeron to the proton,
$f_B (m^2_t)$ is an unknown function of $m_t$ and 
$s_0 \simeq 1$ GeV is a Regge energy scale.
The $\cosh(y/2)$ rapidity dependence 
and the $1/\sqrt[4]{s}$ 
energy dependence follow from the assumed intercept\cite{rossi_77}, 
$\alpha_{M^J_0}(0) \approx 1/2$.  
In contrast to simpler diquark breaking models as
in the Dual Parton Model, 
the multiplicity
of junction also enhanced events is enhanced by a factor of 5/4 in $p+p$,
and the 
strangeness content is also enhanced by a large factor.  
The junction mechanism for baryon number (vs valence quark number)
transport predicts
for the unique possibility of producing $S=-3$ $\Omega^-$ baryons
at midrapidity, as were observed at the SPS in WA97. 
In the Monte Carlo event generator
HIJING/$B\bar{B}$ \cite{OSCAR,Vance:1998vh}, baryon 
junctions are implemented in terms of Y shaped
strings spanning valence quarks.

The junction is a topological knot in the gluon field connecting the
color flux from three quarks into a color singlet state\cite{rossi_77}.
 The intriguing
aspect of junctions is that the conserved baryon number resides in 
the gluon knot and not in the valence quarks\cite{khar_bj96}. 
In a nuclear collision some or 
all of the valence quarks 
may fragment into mesons. However, the gluonic junctions insure that
baryon number is conserved. The understanding of the
dynamics of junction exchange and pair production 
is still rather primitive, but the consistency of the baryon stopping 
power at SPS and now RHIC with  HIJING/$B\bar{B}$ predictions
 suggest that
baryon dynamics at central rapidities may be especially
interesting at RHIC. 
See ref. \cite{Csorgo:2001ng}
 for a discussion of possible novel  junction network
physics that may  lead to
femto-scale buckyball and even CP odd junction network
production in $A+A$.

\subsection{Quenching of Elliptic Flow}

\begin{figure}
\begin{flushleft}
\begin{minipage}{5.5cm}
\includegraphics[width=5.5cm,height=5cm,angle=0]{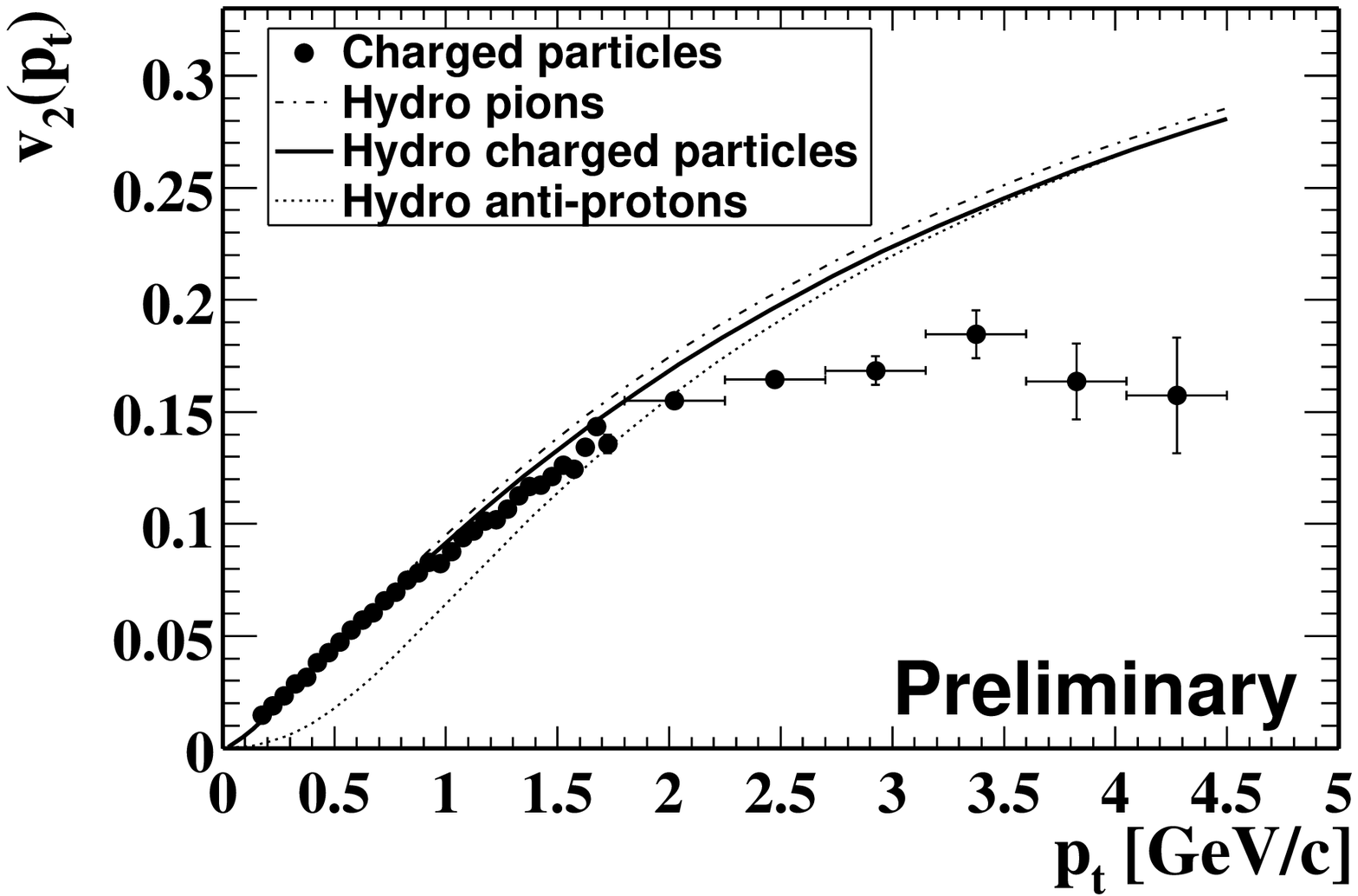}
\caption{Saturation of elliptic flow as measured by 
STAR\protect\cite{Snellings:2001nf}. Curves are the extrapolations
of the hydrodynamic model predictions from \protect\cite{huovinen,Kolb:2001qz}
to high $p_T$.
}
\label{v2hydro}

\end{minipage}
\end{flushleft}
\hspace{\fill}
\vspace{-7.8cm}
\begin{flushright}
\begin{minipage}[t]{5.5cm}
\includegraphics[width=5.5cm,height=5cm,angle=0]{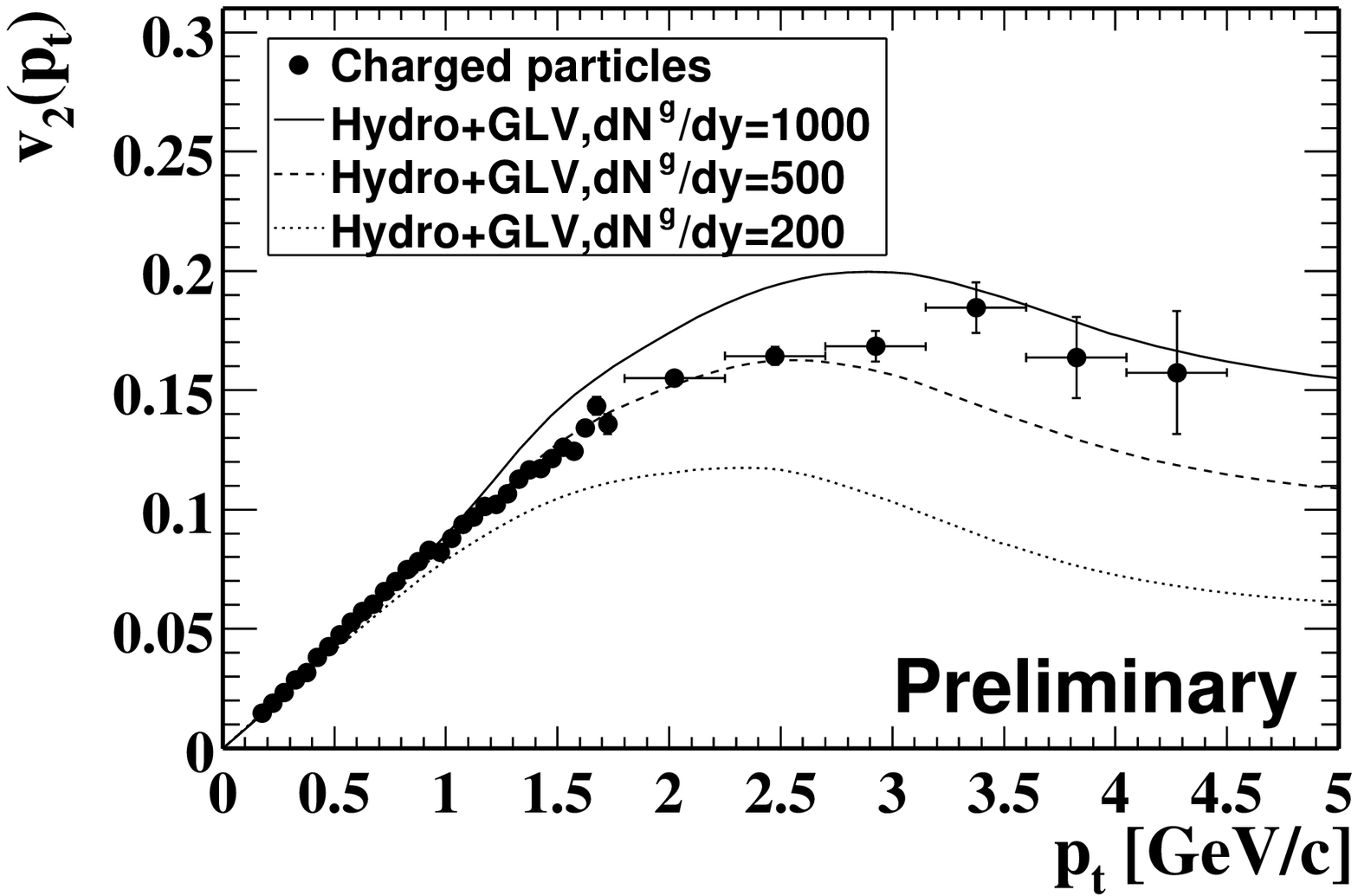}
\caption{Curves show saturation of elliptic flow due to finite energy
loss of partons in a gluon plasma with rapidity density
$dN_g/dy=200,500,1000$ from \cite{gvw}.}
\label{v2pqcd}
\end{minipage}
\end{flushright}
\end{figure}

As seen in Figs.(\ref{pipv2},\ref{v2hydro})
strong elliptic flow
was discovery at RHIC consistent with
hydrodynamic prediction at low transverse momentum $p_T<1$ GeV.
However, the preliminary data from STAR\cite{Snellings:2001nf} 
shows that 
above $p_T>2$ GeV the elliptic flow saturates and 
the azimuthal asymmetry deviates more and more 
from hydrodynamic behavior as seen in
Fig.\ref{v2pqcd}.
This information provides insight into how hydrodynamic behavior
breaks down at high $p_T$ due to the finite energy loss of partons
in the plasma. As shown in detail in \cite{gvw} the saturation pattern 
at high $p_T$ depends
on the energy dependence of the gluon energy loss
 as well as on the geometry
of the plasma density at finite impact parameters. It therefore
provides tomographic information about the density profile
and its evolution in $A+A$. See sec 4.4 for more details.

\subsection{Where did the slowly burning plasma log vanish?}

The last major RHIC result that I highlight here is on
pion interferometry. Relativistic combustion theory\cite{VanHove:1985zy,Gyulassy:1984rq,dirk}
 predicts
that if there were a sufficiently rapid cross over
between the QGP and hadronic phases of ultra-dense matter, then 
a deflagration burn front may appear
between two phases. The main characteristic of that burn front
is its very small velocity in case the entropy density jump across
it is sufficiently large and no high degree of non-equilibrium supercooling
arises.
Even with a smooth cross-over transition, such slowly burning
plasma solutions were shown to exists as long as the width of the transition
region is $(\Delta T_c/T_c<0.08$). The lifetime of 
a Bjorken plasma log is therefore significantly enhanced
$\tau\sim R/v_d$, where $v_d\sim 1/25$ is the small deflagration velocity
in the static 1+1D case.

This characteristic time delay of the hadronization from a QGP state
was suggested
in \cite{pratt,bertsch} to be testable  via pion interferometry.
In ref.\cite{dirk} the  3+1D hydrodynamic equations were
solved to study this plasma ``stall'' phenomenon in detail.

The two pion correlation 
function measures the coincidence probability
$P({\bf p}_1, {\bf p}_2)$ of two (identical) bosons with 
momenta ${\bf p}_1,\, {\bf p}_2$ relative
to the probability of detecting uncorrelated particles
from different events,
\begin{equation}
C_2 ({\bf p}_1, {\bf p}_2) = \frac{P({\bf p}_1, {\bf p}_2)}{P({\bf p}_1)
P({\bf p}_2)}\, .
\end{equation}
If the average 4--momentum is denoted as $K^{\mu} = (p_1^{\mu}
+p_2^{\mu})/2$ and the relative 4--momentum as $q^{\mu}=p_1^{\mu} - p_2^{\mu}$,
then under
the assumption that the particle 
source is chaotic and sufficiently large,
\begin{equation} \label{c2}
C_2 ({\bf p}_1, {\bf p}_2) = 1 + \frac{\left| \frac{1}{(2\pi)^3}
\int_{\Sigma}\, {\rm d}
\Sigma \cdot K\,\, \exp\, [i\, \Sigma \cdot q] \,\, 
f\left(\frac{u \cdot K}{T}\right)
\right|^2}{ E_1 \, \frac{{\rm d}N}{{\rm d}^3 {\bf p}_1} \,\,\, E_2\,
\frac{ {\rm d}N}{{\rm d}^3 {\bf p}_2} }\,\, ,
\end{equation}
where \cite{Cooper:1975qi}
\begin{equation} \label{single}
E\, \frac{{\rm d}N}{{\rm d}^3 {\bf p}} = \frac{1}{(2\pi)^3}
\int_{\Sigma}\, {\rm d}
\Sigma \cdot p \,\, f\left( \frac{u \cdot p}{T}\right)
\end{equation}
is the single inclusive momentum distribution, $f(x)=(e^x-1)^{-1}$, and
$u^{\mu}$ the fluid 4--velocity. The integrals
run over the assumed freeze--out hypersurface. 
In general, that hypersurface is 
represented by a 3--parametric (4--vector) function 
$\Sigma^{\mu}(\zeta,\eta,\phi)$, and
the normal vector on the hypersurface is determined by
\begin{equation} \label{normal}
{\rm d} \Sigma_{\mu} = \epsilon_{\mu \alpha \beta \gamma}\,
\frac{\partial \Sigma^{\alpha}}{\partial \zeta}\, 
\frac{\partial \Sigma^{\beta}}{\partial \eta}\, 
\frac{\partial \Sigma^{\gamma}}{\partial \phi}\, 
{\rm d}\zeta\, {\rm d}\eta\, {\rm d}\phi\,\, ,
\end{equation}
where $\epsilon_{\mu \alpha \beta \gamma}$  is the completely
antisymmetric 4 tensor. For the common isotherm
freeze-out
temperature $T_f$ hypersurface, 
the fluid velocity gerenally varies $u^{\mu} = u^{\mu}(\Sigma)$.

For the Bjorken cylinder geometry, it is useful
to restrict consideration to
particles emitted at midrapidity, $K^z=q^z=0$. Rotational symmetry
around the $z$--axis in central collisions makes it possible
to choose the average
transverse momentum as ${\bf K}_\perp = (K,0,0)$, and consequently,
$C_2(K,q_{\rm \,out},q_{\rm \,side})$ is
a function of three independent variables only.
The so called out and side projections of the relative momenta are
${\bf q}_{\rm \,out}=(q_{\rm \,out},0,0)$, ${\bf q}_{\rm \,side} = 
(0,q_{\rm \,side},0)$. As shown in \cite{pratt,bertsch} the width,
$1/R_{side}$, of the correlation function in $q_{\rm \, side}$ 
is a measure of the transverse 
decoupling or freeze-out radius, 
while the width $1/R_{out}$ of the $q_{\rm \, out}$ correlation function is also sensitive to the duration of hadronization, $\Delta \tau$ 
$$ R_{out}^2\approx R_{side}^2 + v^2 \Delta \tau^2 \;\; .$$
Thus a QGP stall would manifest experimentally in
$R_{out}\gg R_{side}$. In \cite{dirk} it was found that
for possibly realistic parameters, $R_{out}/R_{side}\sim 2-3$, could be observed if a QGP stall occurred.
\begin{figure}
\includegraphics[width=11cm,height=7cm,angle=0]{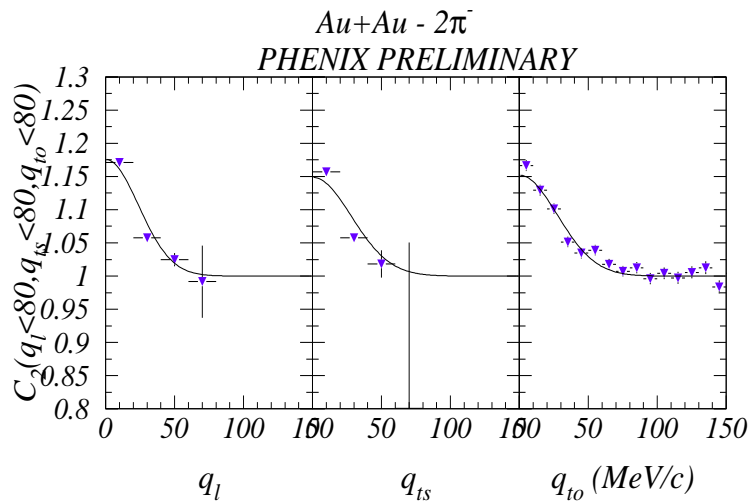}
\caption{Preliminary PHENIX pion interferometry 
data\protect\cite{Johnson:2001zi}
 vs different
projections of the relative momenta. 
Similar preliminary STAR data\cite{harrisqm01} were also shown at QM01.
Unlike predictions\protect\cite{dirk}
there is no hint of the expected stall or time delay of the QGP transition.}
\label{phenixpipi}
\end{figure}

With this ``warm-up'' review of pion interferometry,
we are now ready for the rude awakening from
ideal gedanken considerations with the first splash
of  ``cold'' RHIC pion interferometry data
shown in Fig.(\ref{phenixpipi}). The preliminary PHENIX data show that 
$R_{out}\sim R_{side}$ and even more disturbing all the deduced interferometry
parameters are virtually identical to values seen at the AGS and SPS.
To add insult to injury, it appears that $R_{out} <R_{side}$
for $p_T>0.4 $ GeV.
Preliminary STAR data\cite{harrisqm01} show the same tendency.

Of course scenarios may be invented to ``explain'' the data a postiori,
but if these data are confirmed by further measurements,
then they are indeed surprising and call into question our picture
of the space-time evolution  of $A+A$. That this problem
is not restricted to
idealized hydrodynamics can be seen from the results of 
Ref.\cite{Soff:2001eh}. It was shown that
 $R_{out}>R_{side}$ is also predicted
 in a calculation where the 
entropy jump is small and pion decoupling is dynamically handled via
URQMD. Among the  theoretical questions
that  should now be further investigated
is whether the pion interferometry theory
based on {\em chaotic} ensembles\cite{mg79} is in fact
applicable to $A+A$.
Another question that needs further study 
is whether  the assumed ensemble of initial conditions was 
too restrictive and whether highly inhomogeneous
and turbulent initial conditions apply\cite{turb}.

\section{Jet Quenching: Theory}

Having had a brief tour of some of the interesting new data 
harvested from RHIC during the first round of experiments,
 I turn next  to
the more specific theoretical problem
 of computing the energy loss per unit
length of a fast parton penetrating a finite, expanding quark-gluon plasma.
As I emphasized above, high $p_T$ many body pQCD physics
is a new frontier at RHIC and higher energies. This requires the development
of the  non-abelian analogue of the radiative energy loss theory
familiar from classical E\&M. The interesting new twist is that we have no
external  beams of quarks or gluons and the medium is very thin 
due to the fact that nuclei are tiny. Also the 
formation time physics of Landau-Pomeronchuk-Migdal (LPM) results
in major destructive interference effects that must be taken into account.
Work on this problem over the past five years has advanced considerably
but many open problems remain.

I will only highlight only one of those direction,
 namely the opacity expansion
reaction operator method that we developed in refs.\cite{glv}.
The reader is referred to BDMS\cite{bdms8}, Z\cite{Zakh00}, and U\cite{urs00}
for alternative methods and approximations.

In ref.\cite{Gyulassy:1994hr} we proposed a simple model to study induced gluon
radiation due to multiple elastic scattering of a high energy jet propagating in a locally color neutral amorphous plasma.

\subsection{GLV Formalism}

\begin{figure}
\begin{center}
\includegraphics[width=6cm,height=3cm,angle=0]{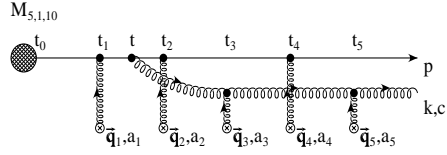}
\caption{Induced radiation amplitude\protect\cite{glv1} contributing
to fifth order and higher order in the opacity expansion of QCD energy loss
in the GW model\protect\cite{Gyulassy:1994hr}. The crosses denote static color screened Yukawa interactions on a scale $\mu$. The blob is the initial hard jet amplitude without final state interactions.}
\label{psmq5xfig}
\end{center}
\end{figure}

\begin{figure}
\begin{center}
%
\begin{center}
\begin{minipage}{8cm}
\hspace*{-3.5cm}
\includegraphics[width=11cm,angle=270]{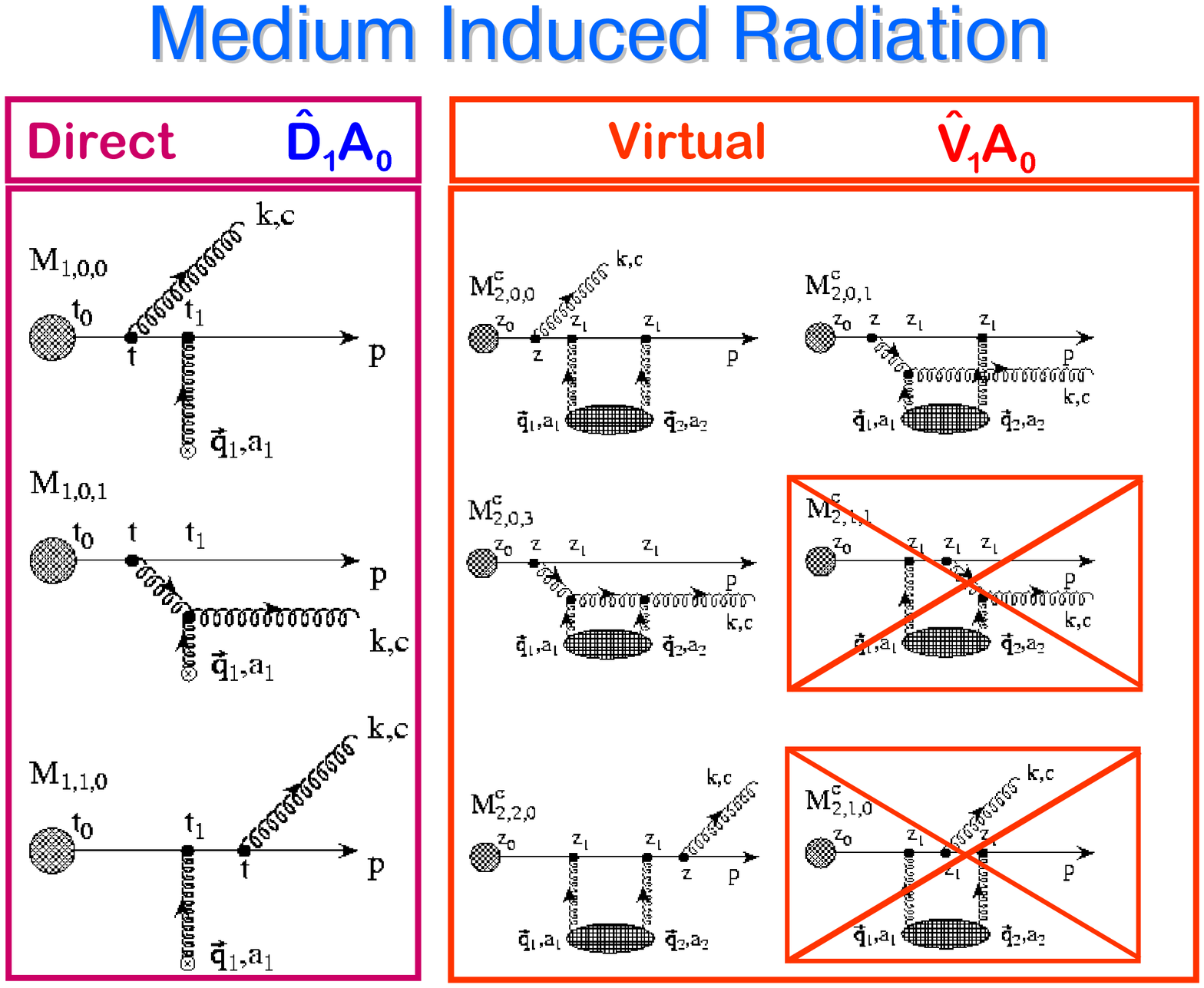}
\end{minipage}
\end{center}
\caption{Three first order (singe Born) direct and four surviving
(double Born)
virtual or contact amplitudes\protect\cite{bdms8} from which
the 
$\hat{D}_n$ and $\hat{V}_n$ components
of the reaction operator in eq.(\protect\ref{reacop})
are derived in \protect\cite{glv}.}
\label{dandv}
\end{center}
\label{m5x}
\end{figure}

In \cite{glv1} we developed a systematic graphical method to compute
medium induced  gluon radiation amplitudes as shown
in Fig.\ref{psmq5xfig}.
The exponential growth of the number of graphs with the number of interactions
makes it very tedious to go beyond order three.
In GLV\cite{glv} we overcame the combinatorial problem by developing a new 
algebraic operator technique to solve 
the inclusive radiated gluon distributions recursively.
The first step is to compute
the three direct (single Born) and four surviving 
virtual (contact double Born) diagrams
shown in Fig.\ref{dandv}.

\begin{figure}
\begin{center}
\hspace{-1.5cm}
\includegraphics[width=10cm,angle=270]{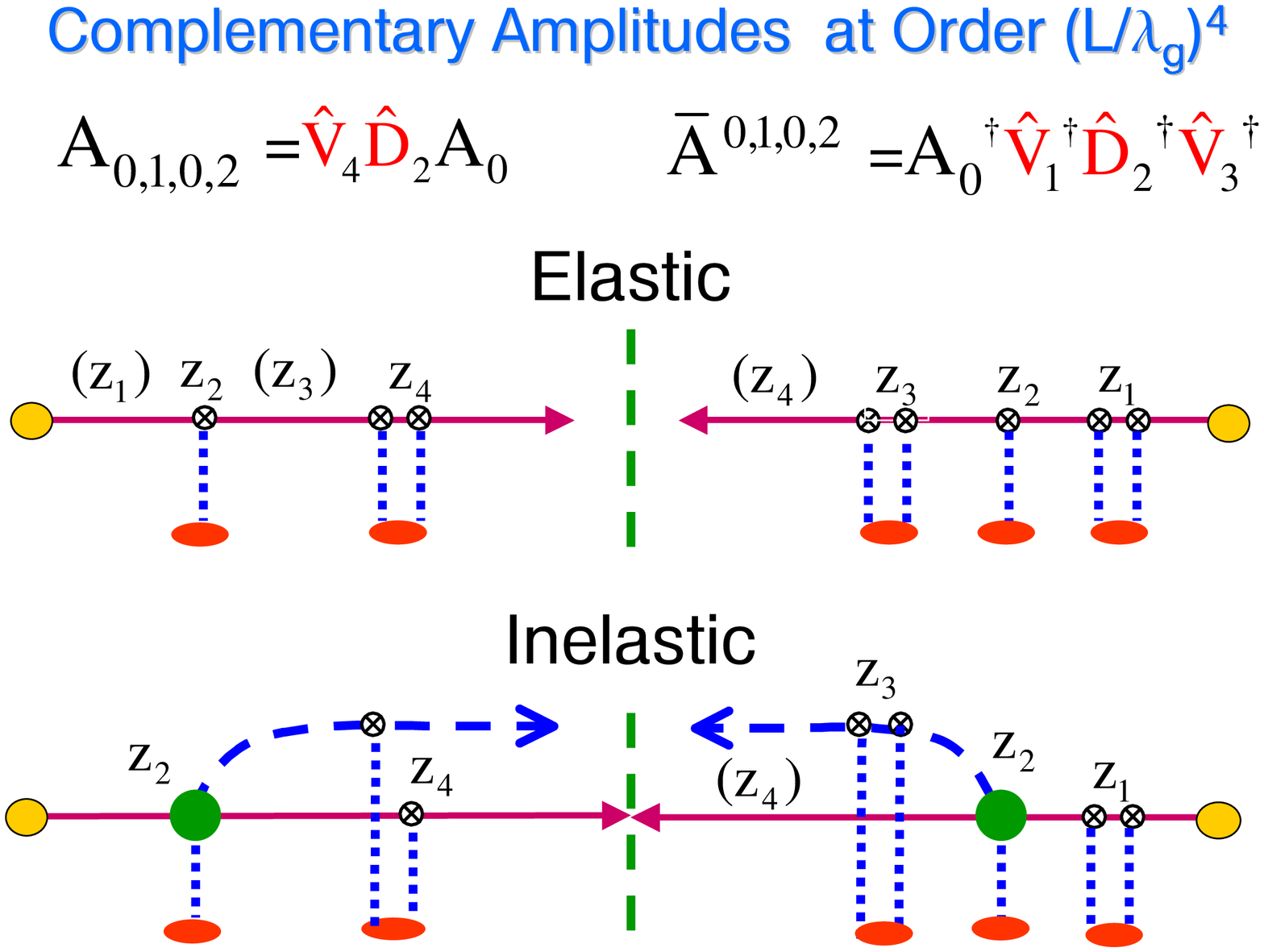}
\vspace{-1cm}
\caption{Example of graphs   constructed via $\hat{D}_i,\hat{V}_i$
that contribute to the 4th order in opacity in elastic and inclusive
inelastic final state interactions. The longitudinal depth of active
scattering centers are denoted by $z_i$ and inactive (created with 
$\hat{1}_i$)
 by $(z_i)$. The form of $\hat{D}_i,\hat{V}_i$ depend on the process
type but the tensorial bookkeeping of partial sums of amplitudes is 
the same.}
\label{4thorder}
\end{center}
%
\end{figure}
\renewcommand{\vec}[1]{{\bf #1}}

\newcommand{\vAi}{{\cal A}_{i_1\cdots i_n}}
\newcommand{\vAim}{{\cal A}_{i_1\cdots i_{n-1}}}
\newcommand{\vAbi}{\bar{\cal A}^{i_1\cdots i_n}}
\newcommand{\vAbim}{\bar{\cal A}^{i_1\cdots i_{n-1}}}
\newcommand{\htS}{\hat{S}}
\newcommand{\htR}{\hat{R}}
\newcommand{\htB}{\hat{B}}
\newcommand{\htD}{\hat{D}}
\newcommand{\htV}{\hat{V}}
\newcommand{\cT}{{\cal T}}
\newcommand{\cM}{{\cal M}}
\newcommand{\cMs}{{\cal M}^*}
\newcommand{\vk}{{\vec k}}
\newcommand{\vK}{{\vec K}}
\newcommand{\vb}{{\vec b}}
\newcommand{{\vq}}{{\vec q}}
\newcommand{\vQ}{{\vec Q}}
\newcommand{\tr}{{{\rm Tr}}}
\newcommand{\half}{\frac{1}{2}}
For scattering off of $n$ scattering centers located at depths $z_i$ in a
transverse homogeneous medium of large area ($(\mu R)^2\gg 1$),
we can write the inclusive radiated gluon spectrum, $P_n({\bf k},c)$,
 as a sum
over products of partial sums of  amplitudes and complementary
complex conjugate amplitudes. Every term in the sum contributes
to the same $O(g^{2n})$. The average value
of $n$ is referred to as the opacity of the medium.
The partial sums of diagrams at order $n$ in such and opacity expansion
can be conveniently expressed in a tensor notation
and  constructed
by repeated operations of, $\hat{\bf 1},\htD_i$, or $\htV_i$
corresponding to no, direct, or virtual interactions at 
scattering center $i$
$$\vAi(x,{\bf k},c)=\prod_{m=1}^n
\left( \delta_{0,i_m} + \delta_{1,i_m} \htD_m + \delta_{2,i_m} 
\htV_m
\right) G_0(x,{\bf k},c)
\; \; .$$
Here $G_0$ is the initial hard $q+g$ color matrix amplitude.
In the inclusive probability each class contracts with
a unique {{ complementary}} class
$$P_n(x,{\bf k})=\vAbi(c)\vAi(c)
$$
with the complementary class constructed as
$$\vAbi(x,{\bf k},c) \equiv 
G_0^\dagger
(x,{\bf k},c) \prod_{m=1}^n
\left( \delta_{0,i_m} \hat{V}_m^\dagger + 
\delta_{1,i_m} \hat{D}_m^\dagger 
+ \delta_{2,i_m} 
\right)$$ 
Fig.(\ref{4thorder}) shows an example of how this
formalism works at 4th order in opacity for elastic and inelastic
inclusive distributions.

Direct  interactions 
enlarge  rank $n-1$ class elements as follows:
$$
 \htD_n \vAim(x,{\bf k},c) \equiv  (a_n + \htS_n + \htB_n) 
\vAim(x,{\bf k},c)
$$
$$= {a_n} \vAim(x,{\bf k},c) + 
e^{i(\omega_0-\omega_n)z_n } 
\vAim(x,{\bf k}- {{\bf q}_{n}},{[c,a_n]}) -$$
$$  {\left(-\half \,\right )^{N_v(\vAim)}} {\bf B}_n \, 
e^{i \omega_0 z_n} {[c,a_n]} T_{el}(\vAim) $$
where ${\bf B}_n \, = {\bf H}-{\bf C}_n=\vk/\vk^2-(\vk-\vq_n)/(\vk-\vq_n)^2)$
 is the so-called Bertsch-Gunion amplitude
for producing a gluon with transverse momentum $\vk$
in an isolated single collision
with scattering center $n$. The momentum transfer to the jet is
$\vq_n$.
The notation $\omega_n=(\vk-\vq_n)^2/2\omega$, 
for a gluon with energy $\omega$
and $a_n$ is the color matrix in the $d_R$ dimensional
representation of the jet with color Casimir $C_R$.
$N_v=\sum^{n-1}_{m=1}\delta_{i_m,2}$ 
counts the number of virtual interactions
in $\vAim$. 

Unitarity (virtual forward scattering) corrections 
to the direct processes
involve the sum of four double born contact diagrams in Fig.(\ref{dandv})
that 
enlarge  rank $n-1$ classes as follows:
\begin{equation}\htV_n  =  -\half(C_A+C_R) - a_n \htS_n- a_n\htB_n
= -a_n \htD_n - \half(C_A-C_R)\label{key}
\end{equation}
This {\em {key}} operator relation between direct and virtual
insertions that we discovered  in \cite{glv}
makes it possible to solve the problem algebraically.

The tensor classification of classes of diagrams
makes it possible to construct the distribution of radiated gluons
in the case of $n$ interactions,
$P_n$, recursively  from lower rank (opacity) classes
via a {``reaction''} operator
\begin{equation}
P_n=\bar{\cal A}^{i_1\cdots i_{n-1}}\htR_n {\cal A}_{i_1\cdots i_{n-1}}
\;, \qquad \htR_n
= \hat{D}_n^\dagger
\hat{D}_n+\hat{V}_n+\hat{V}_n^\dagger
\label{reacop}
\end{equation}
Using the key identity (\ref{key}), the reaction matrix simplifies to
$${\htR_n=  (\htD_{n}-a_{n})^\dagger  (\htD_{n}-a_{n}) - C_A}
{=(\htS_{n}+\htB_{n})^\dagger  (\htS_{n}+\htB_{n}) -C_A}$$

The next major simplification
occurs because both $\htS$ and $\htB$ involve the same
gluon color rotation through  $if^{ca_{n} d}$. This reduces the
color algebra to multiplicative Casimir factors
$$
\vAbim(\htS_{n}^\dagger \htS_{n}-C_A)\vAim 
$$
$$= C_A\left( P_{n-1}({\bf k}-{\bf q}_{n})
            - P_{n-1}({\bf k})\right)
= C_A\left( e^{i{\bf q}_{n}\cdot\hat{\bf b}}-1 \right) P_{n-1}({\bf k})$$
$$\vAbim\htB_{n}^\dagger \htB_{n}\vAim =0 $$
$$2 {\bf Re}\, \vAbim\htB_{n}^\dagger \htS_{n}\vAim
= -2 C_A\, {\bf B}_n\cdot \left({\rm Re}\; e^{-i\omega_nz_n} 
e^{i{\bf q_n}\cdot\hat{\bf b}} {\bf I}_{n-1}\right)$$
 ${\bf I}_n$ obeys
 a recursion relation from which the  inclusive radiation
probability  is found to obey the soluble
recursion  relation
$$ P_{n}({\bf k})= C_A
(P_{n-1}({\bf k}-{\bf q}_{n}) - P_{n-1}
({\bf k)}) 
-2 C_A\, {\bf B}_n\cdot \left( {\bf Re}\; e^{-i\omega_nz_n} 
e^{i{\bf q_n}\cdot\hat{\bf b}} {\bf I}_{n-1} \right)$$
$$ + \delta_{n,1} C_A C_R |{\bf B}_1|^2$$
where $\hat{\bf b}=i\nabla_{\bf k}$ is the transverse momentum
shift operator. The initial condition for this recursion relation
is the initial hard vertex radiation amplitude without final state
interactions that is given by
$P_0=C_R \, {\bf H}^2={C_R}/{{\bf k}_\perp^2} $.

The complete solution to the problem can therefore
be expressed in closed form as
$$P_{n}({\bf k})= -2{C_R C_A^n} \, {\bf Re}
\sum_{i=1}^n 
\left\{\prod_{j=i+1}^n( e^{i{\bf q}_{j}\cdot \hat{\bf b}} - 1)  
\right\}
{\otimes} {\bf B}_{i} \cdot\; 
e^{i{\bf q}_{i}\cdot \hat{\bf b}}   e^{-i\omega_0 z_i} \times
$$
$$\left\{
    \prod_{m=1}^{i-1}(e^{i(\omega_0-\omega_m)z_m}
     e^{i{\bf q}_{m}\cdot \hat{\bf b}} -1)
  \right\}
{\otimes}   {\bf H}(e^{i\omega_0 z_1}-e^{i\omega_0 z_0})$$
This expression
can be averaged over any  spatial distribution of interaction centers,
$z_i$ as well as any $z_i$ dependent momentum transfers $q_n$.
This form is thus  
ideally suitable for Monte Carlo implementation
for arbitrary ${\bf q}_i, z_i$
medium ensemble averages.
 
\subsection{Nonabelian Energy Loss at Finite Opacity}

The first application\cite{glv} of our general solution to the energy loss
problem was to calculate numerically the total
radiated energy loss as a function of jet energy $E$, plasma depth, $L$,
and infrared screening scale, $\mu$.
In the absence of a medium, the gluon  radiation
associated with a spin $\half$ parton jet is distributed as
\begin{equation}  
x\frac{dN^{(0)}}{dx\, d {\bf k}^2_\perp}= 
 \frac{C_R \alpha_s}{\pi}
\left( 1-x+\frac{x^2}{2}\right)
\frac{1}{{\bf k}^2_\perp} \;\;, 
\label{hdist}
\end{equation}
where 
$x=k^+/E^+ \approx \omega/E$, and $C_R$ is the Casimir 
of the (spin $1/2$) jet in the $d_R$ dimensional color representation.
The differential energy distribution outside a cone defined by
${\bf k}^2_{\perp}>\mu^2$ is given by
\begin{equation} 
\frac{dI^{(0)}}{dx} = \frac{2 C_R \alpha_s}{\pi} 
\left( 1-x+\frac{x^2}{2}\right)
\, E \, \log \frac{|{\bf k}_\perp|_{\rm max}}{\mu} \;,    
\label{di0}
\end{equation} 
where the upper kinematic limit is
$\quad {\bf k}^2_{\perp \, \max}=\min\, [4E^2x^2,4E^2x(1-x)]\;.
$. 
The energy loss outside the cone in the vacuum is then given by
\begin{equation}
\Delta E^{(0)}=\frac{4C_R\alpha_s}{3\pi}\, E \,
\log \frac{E}{\mu}
\label{de0}
\end{equation}
While this overestimates the 
radiative  energy loss in the vacuum (self-quenching), it is important to note
that $\Delta E^{(0)}/E \sim 50\%$ is typically much larger
than the medium induced energy loss.

Averaging over the momentum transfer ${\bf q_{1\perp}}$ via the color
Yukawa potential leads to a very simple first order opacity 
result for  the $x\ll 1$ gluon double differential distribution
\begin{eqnarray}
x\frac{dN^{(1)}}{dx\, d {\bf k}^2_\perp}&=& 
x\frac{dN^{(0)}}{dx\, d {\bf k}^2_\perp} 
\, \frac{L}{\lambda_g}  \int_0^{q_{\max}^2} d^2{\bf q}_{1\perp} \, 
\frac{ \mu_{eff}^2 }{\pi ({\bf q}_{1\perp}^2 + \mu^2)^2 }
\nonumber \\[.5ex]
&& \qquad \qquad \qquad \frac{ 2\,{\bf k}_\perp \cdot {\bf q}_{1\perp}
  ({\bf k} - {\bf q}_1)_\perp^2  L^2}
{16x^2E^2 +({\bf k} - {\bf q}_1)_\perp^4  L^2 } \;\;.          
\label{dnx1}
\end{eqnarray}
where the opacity factor $L/\lambda_g=N\sigma_{el}^{(g)}/A_\perp$ 
arises from the sum over the $N$ distinct
targets. Note that the radiated gluon mean free path
$\lambda_g=(C_A/C_R)\lambda$ appears rather than the jet mean free path. 
The upper kinematic bound on the momentum transfer
 $q^2_{\rm max}= s/4 \simeq 3 E \mu$, ($1/\mu_{eff}^2=1/\mu^2-
1/(\mu^2+q_{\max}^2)$). For SPS and RHIC energies,
this finite limit cannot be ignored as we show below.

The  second order contribution in opacity
involving the sum of  $7^2$ direct and $2\times 86$ virtual 
and results  in \cite{glv}  
\begin{eqnarray}
&& P^{(2)} \propto 
C_R C_A^2 d_R \, \left[ \, 
2\,{\bf C}_1 \cdot {\bf B}_1 \,
\left( 1- \cos ( \omega_1 \Delta z_1 )\right) \right.  
\nonumber \\[.8ex]
&&\quad +\, 2\, {\bf C}_2 \cdot {\bf B}_2 \,
\left( \cos ( \omega_2 \Delta z_2 ) - 
\cos ( \omega_2 (\Delta z_1 + \Delta z_2 )\right) 
\nonumber \\[.8ex]
&&\quad- \,2\, {\bf C}_{(12)} \cdot {\bf B}_2 \,
\left( \cos ( \omega_2 \Delta z_2 )
 - \cos ( \omega_{(12)} \Delta z_1 + \omega_2 \Delta z_2 )\right) 
\nonumber \\[.8ex]
&&\left. \quad -   \,2\, {\bf C}_{(12)} \cdot {\bf B}_{2(12)} \,
\left( 1- \cos ( \omega_{(12)} \Delta z_1 )\right) \, \right]\;, 
\label{secord}
\end{eqnarray} 
where with ${\bf C}_{(mn)}$ and $\omega_{(mn)}$ obtained from    
${\bf H}$ and $\omega_0$
through the substitution ${\bf k}_\perp \Rightarrow 
{\bf k}_\perp - {\bf q}_{\perp m}-{\bf q}_{\perp n} $ and 
${\bf B}_{m(nl)} \equiv  {\bf C}_m - {\bf C}_{(nl)}$\cite{glv}.

\begin{figure}
\begin{center}
\includegraphics[width=7cm,angle=270]{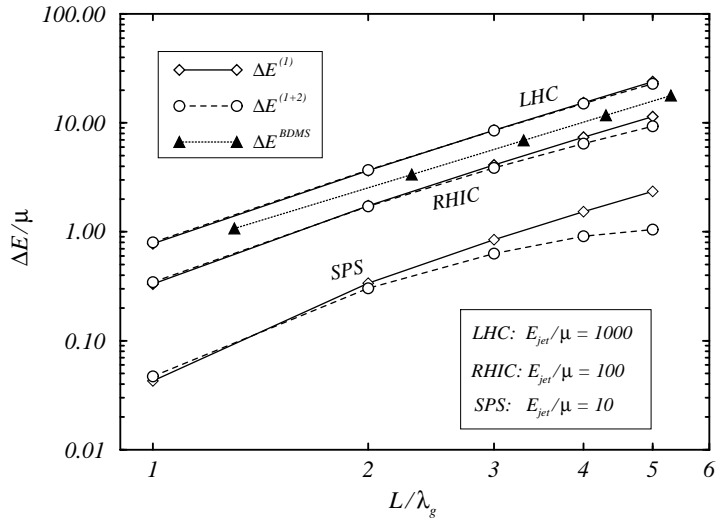}
\caption{\small The GLV radiated energy loss\protect\cite{glv}
        of a quark jet
        with energy $E_{jet}=5,50,500$~GeV  (at SPS, RHIC, LHC) 
        is plotted as a function of 
        the opacity $L/\lambda_g$.  ($\lambda_g=1$~fm, $\mu=0.5$~GeV).
        Solid curves show first order, while dashed curves show
        results up to second order in opacity. 
        The asymptotic energy loss (solid triangles) 
         of BDMS\protect\cite{bdms8}
        is shown for comparison. The energy dependence of GLV suppressing
radiative energy loss of low energy jets explains
why no jet quenching was observed at the SPS (see Fig.\protect\ref{figlet2})
.} 
\label{psglv3}
\end{center}
\end{figure}

Numerical results comparing the first three orders in opacity corrections
to the hard distribution Eq.~(\ref{hdist}) were presented
in \cite{glv}.
To illustrate the result consider a 
quark jet in a medium with $\lambda_g=1$~fm, a screening
scale $\mu=0.5$~GeV and $\alpha_s=0.3$ .  
The total radiative energy loss could be  expressed as
\begin{equation}
\Delta E^{(1)}=\frac{C_R\alpha_s}{N(E)}\, 
\frac{L^2\mu^2}{\lambda_g} \,\log \frac{E}{\mu}  \;\;,
\label{de1}
\end{equation}
with $N(\infty)=4 \log(E/\mu)/\tilde{v}$ 
if the kinematic bounds were ignored as 
in the approximations of ref.\cite{bdms8}. 
We found  that finite kinematic constraints and the form 
of the first order result
cause $N(E)$ to deviate considerably
from the asymptotic value for all energies accessible in the RHIC range.
Together with the logarithmic
dependence on energy, these kinematic effects suppress greatly
the energy loss at lower (SPS) energies as seen in 
Fig.\ref{psglv3}.
This is in sharp contrast to the approximately energy independent result
in BDMS-ZW where the finite kinematic bounds were neglected
because only the asymptotic limits were considered.
  Another remarkable  result  demonstrated numerically is
that the  second and third order  contributions
to  the integrated energy loss remains surprisingly 
small in the physical range of  nuclear 
opacities $L/\lambda_g\sim 5$.
 The rapid convergence of the opacity expansion even for realistic
opacities results from the fact that the effective
expansion parameter is actually the product of the opacity and 
the gluon formation probability $L\mu^2/2xE$.
The leading quadratic dependence
of the energy loss on nuclear thickness discovered in BDMS\cite{bdms8}
therefore already emerges from the dominant
first order term in the opacity expansion.

At SPS energies kinematic effects
suppress greatly the energy loss relative to BDMS.
Our estimates provide a natural explanation for the absence of
jet quenching in $Pb+Pb$ at 160~AGeV observed by WA98.
  At RHIC energies, on the other hand, a significant
nonlinear (in $A$) pattern of suppression of high $p_\perp$
hadrons relative to scaled $pp$ data is predicted.

\subsection{The Opacity of the QGP at RHIC}

\begin{figure}[htb]
\vspace*{-1.2cm}
\begin{minipage}[t]{5.5cm}
\includegraphics[width=6.0cm]{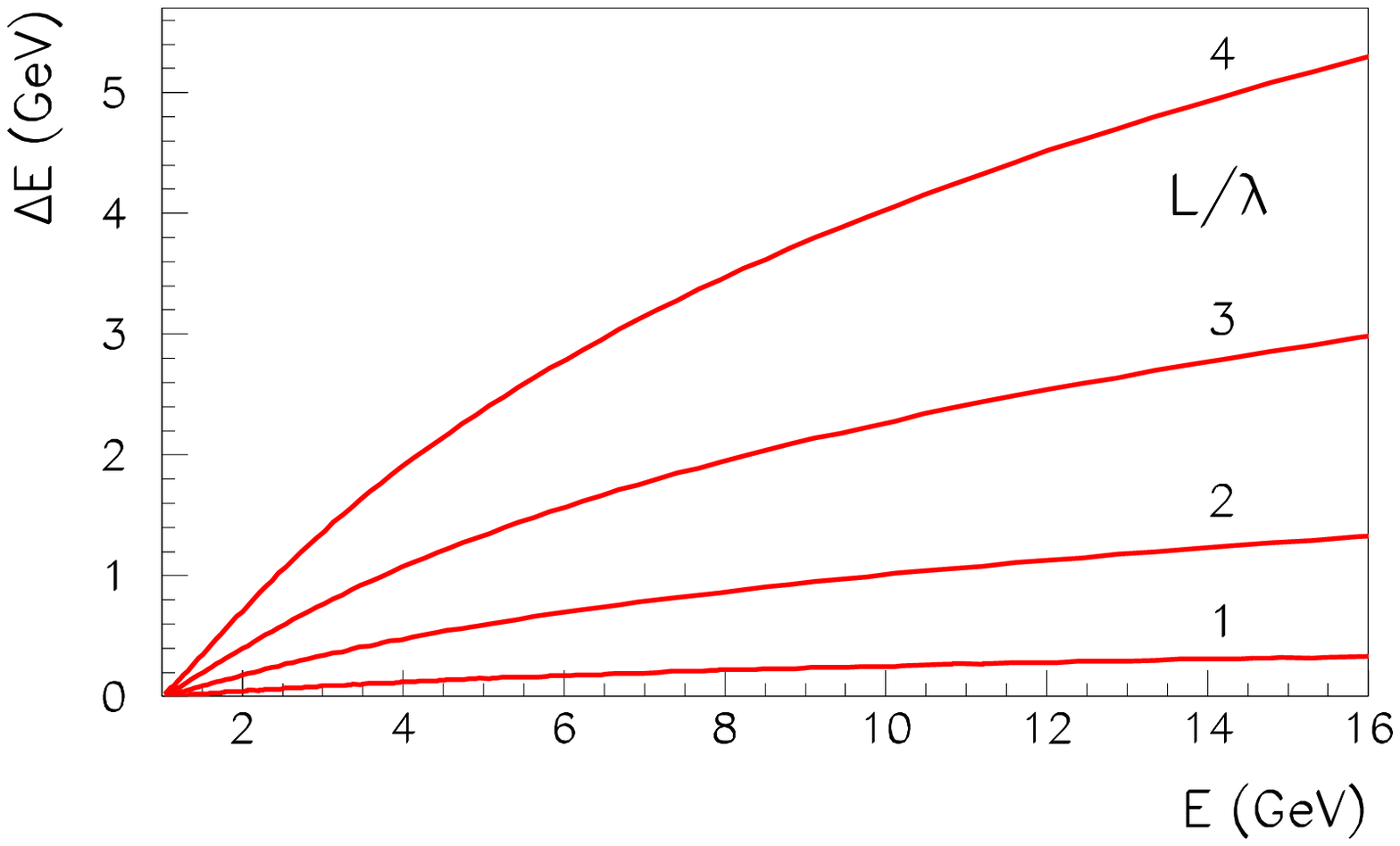}
\vspace{-3cm}
\caption{
Non-abelian energy loss of a gluon jet
calculated in the GLV picture~\cite{glv}.
}
\label{abse}
\end{minipage}
\hspace{\fill}
\begin{minipage}[t]{5.5cm}
\includegraphics[width=6.0cm]{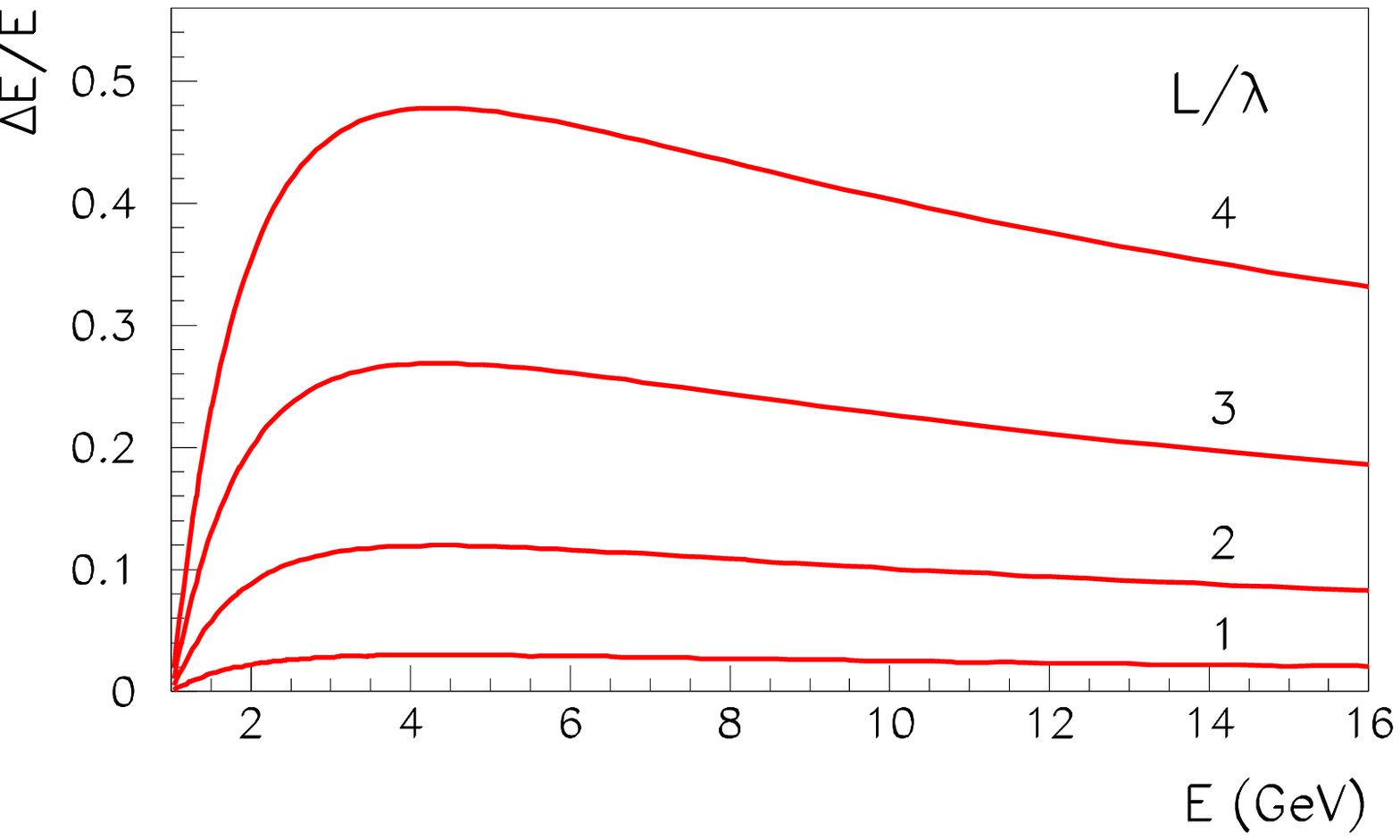}
\vspace{-30mm}
\caption{
The relative energy loss ($\Delta E/E$) is approximately constant
 at medium energy, $2 \leq E \leq 10$ GeV.
}
\label{rele}
\end{minipage}
\end{figure}

As a second a application of the GLV energy loss, in ref.\cite{levaiqm01}
we computed the quenched pQCD distribution of high $p_T$ hadrons
as a function of the effective static plasma opacity, $L/\lambda_g$.
In Figs.(\ref{abse},\ref{rele}), the jet energy dependence of the
GLV energy loss for gluons is shown. The most important feature
to note is that $\Delta E_{GLV}/E$ is approximately constant in the
energy range accessible at RHIC.
\vspace{-1.5cm}

In order to compute the pion spectrum, note
that jet quenching reduces  the energy
of the jet before fragmentation. We concentrate on 
mid-rapidity ($y_{\rm cm}=0$),
where the jet transverse
momentum before fragmentation is shifted by the energy loss
as in ~\cite{WaHu97},
$p_{\rm c}^*(L/\lambda) = p_{\rm c} - \Delta E(E,L)$. This
shifts the $z_{\rm c}$ parameter in the fragmentation function
of the integrand (\ref{fullaa})
to $z_{\rm c}^* = z_{\rm c} /(1-\Delta E/p_{\rm c})$.

The invariant cross section of hadron
production in central $A+A$ collision is then given by\cite{kapiquench}
\begin{eqnarray}
\label{fullaa}
E_{\rm h}\frac{{\rm d}\sigma_{\rm h}^{\rm AA}}{{\rm d}^3p} 
        &=&\int {\rm d}^2 b \  {\rm d}^2 r\ t_{\rm A}({\vec b}) 
         t_{\rm B}({\vec b} - {\vec r})
        \sum_{\rm abcd}\!
        \int\!\!{\rm d}x_{\rm a} {\rm d}x_{\rm b} {\rm d}z_{\rm c}
        {\rm d}^2k_{\perp,{\rm a}} {\rm d}^2k_{\perp,{\rm b}} \cdot
        \nonumber \\
        && \ \ \ \  f_{\rm a/A}(x_{\rm a},k_{\perp,{\rm a}}({\vec b}),Q^2) 
         f_{\rm b/A}(x_{\rm b},k_{\perp,{\rm b}}({\vec b} - {\vec r}),Q^2)\
         \frac{{\rm d}\sigma}{{\rm d}{\hat t}} \nonumber\\
&&\;\; \;\;\;\frac{z^*_{\rm c}}{z_{\rm c}}
   \frac{D_{\rm h/c}(z^*_{\rm c},{\widehat Q}^2)}{\pi z_{\rm c}^2} \,
    {\hat s} \delta({\hat s} + {\hat t} + {\hat u}) \,\, ,
\end{eqnarray}
where upper limit of the impact parameter integral is
 $b_{\rm max}= 4.7$ fm for 10 \% central Au+Au collisions.
Here $t_{\rm A}(b)$ is the usual (Glauber) thickness function.
The factor $z^*_{\rm c}/z_{\rm c}$ appears because of  the in-medium
modification of the fragmentation function~\cite{WaHu97}.
Thus, the invariant cross section
(\ref{fullaa})  depends on the average
opacity or collision number, ${\bar n} = L/\lambda_{\rm g}$.
\begin{figure}[htb]
\vspace*{-0.8cm}
\begin{minipage}[t]{55mm}
\includegraphics[width=60mm]{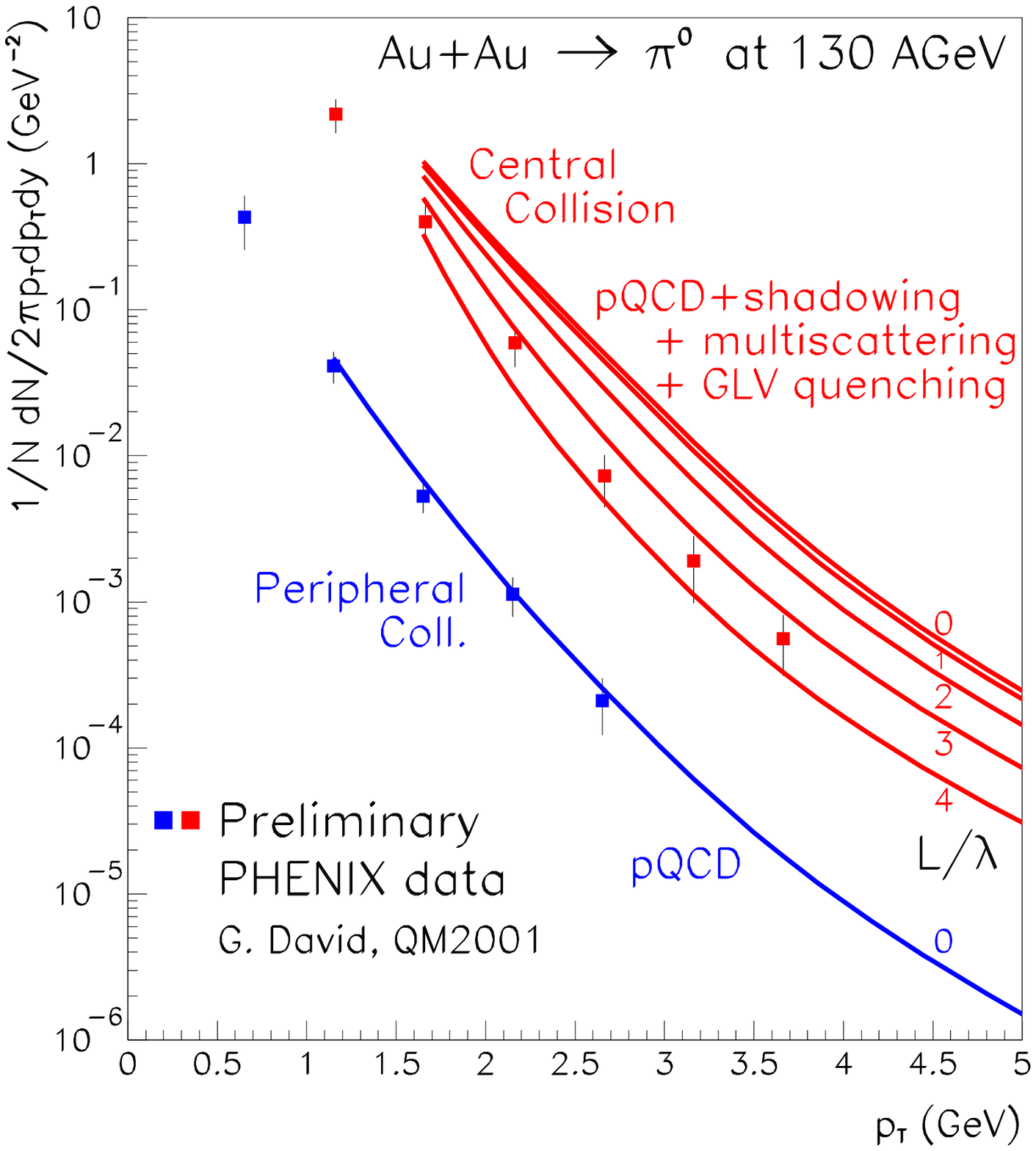}
\vspace{-6mm}
\caption{
Pion production in Au+Au collision
including jet quenching with opacity $L/\lambda=1,2,3,4$.
Preliminary QM01 PHENIX data shown (see updated data from
\protect\cite{davidqm01}
in Fig.\protect\ref{fig:plot_pt_ua1})}
\label{abspi0}
\end{minipage}
\hspace{\fill}
\begin{minipage}[t]{55mm}
\includegraphics[width=60mm]{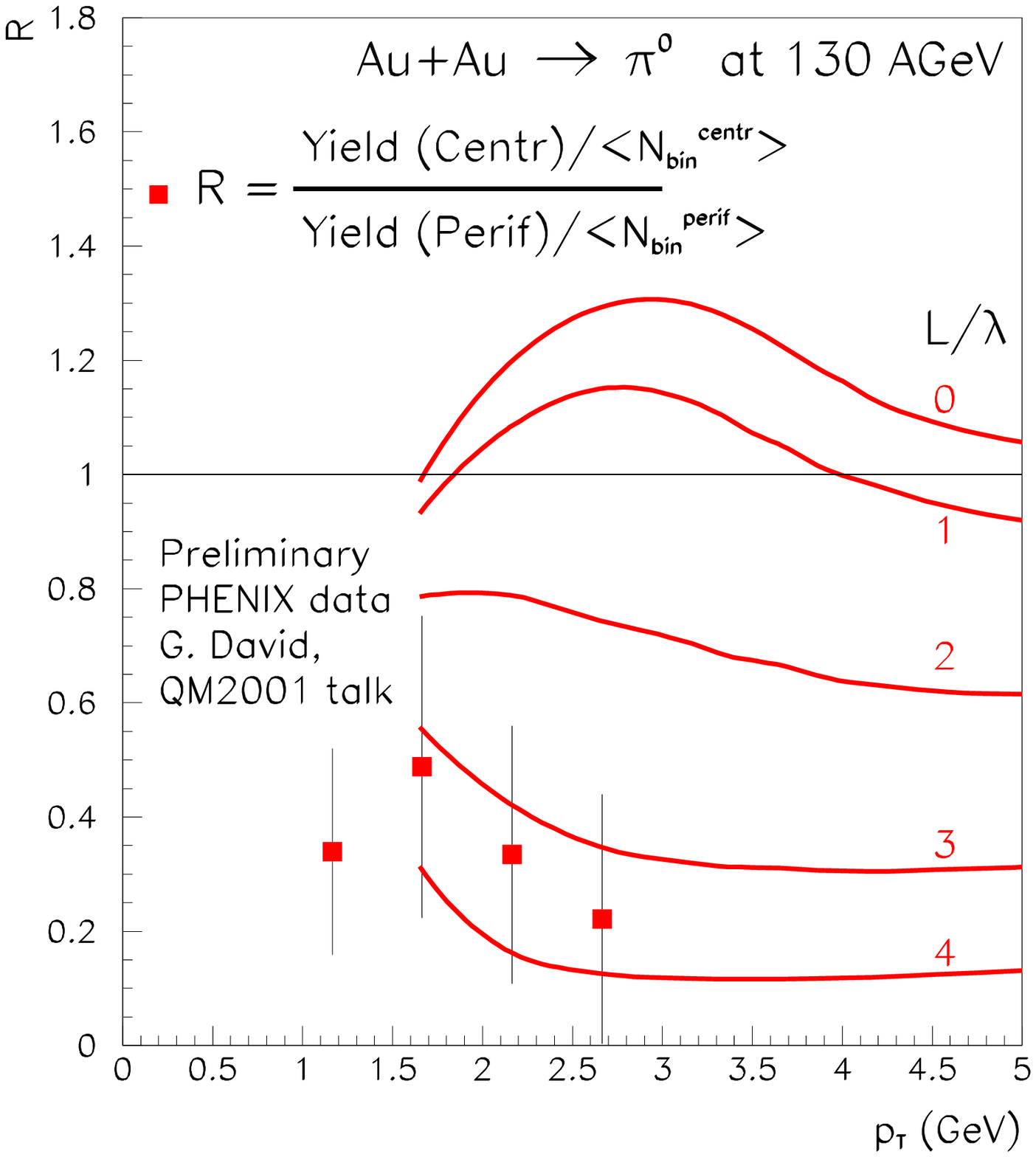}
\vspace{-6mm}
\caption{The ratio of the central to
the peripheral pion yields (normalized by the number of binary collisions,
857 and 5.5). (Note updated data are shown  in Fig.\protect\ref{fig:r-rhic})
}
\label{relpi0}
\end{minipage}
\end{figure}
The calculated spectra for pions  are displayed
for ${\bar n}=0,1,2,3,4$ in Fig. \ref{abspi0}. Fig. \ref{relpi0}. shows
their ratios to the non-quenched spectra at ${\bar n}=0$.
We note that in contrast to previous energy independent
estimates for the energy loss, the GLV energy-dependent
energy loss leads to constant suppression of the high $p_T$
domain in agreement with the preliminary data.
The  peripheral collisions are consistent with a rather small opacity
in contrast to central collisions, as expected.
 The ratio of central to peripheral PHENIX~\cite{davidqm01} data
 from QM01 shown in Fig. \ref{relpi0}
clearly reveals that jet quenching at RHIC overcomes the Cronin
enhancement at zero (final state) opacity. This is in stark contrast to
data at SPS energies, where WA98 found no evidence for quenching
in $Pb+Pb$ at 160 AGeV but a factor of two Cronin enhancement
as discussed before. 

Figs. \ref{abspi0},\ref{relpi0} indicate that an
effective static plasma opacity $L/\lambda = 3-4$ 
is sufficient to  reproduce the preliminary jet quenching pattern observed
at RHIC. In ref.\cite{wangqm01} it was shown that 
 a rather small constant $dE/dx\approx
0.25$ GeV/fm was also found to be consistent with the data.
However, it is important to emphasize that these effective {\em static}
plasma opacities and parameters hide the underlying rapid
dilution of the plasma due to expansion.
The GLV formalism including the kinematic constraints
at first order has been further generalized to include effects of 
expansion in~\cite{gvw}. 
It was found in~\cite{gvw} that  the inclusion of longitudinal
expansion modifies the static plasma results in such a way that the 
moderate static plasma opacity actually
 implies that the produced  mini-jet 
plasma rapidity density may have reached $dN_g/dy\sim 500$.

\subsection{Jet Tomography from Quenched Elliptic Flow}

So far we have not included the dilution effect
of expansion on the energy loss. The generalization of GLV to the case
of expanding plasmas is\cite{gvw} 
\begin{eqnarray} 
\frac{dI_{GLV}}{dx} &=& 
\frac{9 C_R   E}{\pi^2} 
\int\limits^\infty_{z_0} \!d z\,\rho(z)\!\!\!
\int\limits^{|{\bf k}|_{\rm max} } \!\!\! d^2{\bf k}\, \alpha_s\!\!\!
\int\limits^{|{\bf q}|_{\rm max} }  \!\!\!
\frac{d^2{\bf q}\; \alpha_s^2}{({\bf q}^2+\mu(z)^2)^2} \,
\nonumber \\[0.ex]
&\;& 
\frac{{\bf k}\cdot{\bf q}}{{\bf k}^2({\bf k}-{\bf q})^2} 
\left[ 1
-   \cos \left (\,\frac{({\bf k}-{\bf q})^2}{2 x E}(z-z_0) \right)
\right] \; .
\label{ndif1}\end{eqnarray} 
where $\rho(z)$ is the plasma density at time $z$ along the jet
path at position $z$ from the production point and where
the screening scale $\mu(z)$ may also depend on time.

Consider a density evolution of the form as in \cite{bdms8}, 
\begin{equation}
\rho(z)=\rho_0\left(\frac{z_0}{z}\right)^\alpha \theta(L-z)\;,
\label{rhoex}\end{equation}
where $\alpha=0$ corresponds to a static uniform medium of thickness $L$,
while $\alpha=1$ to Bjorken 1+1D longitudinal expansion
transverse to the jet propagation axis.  

Analytic expressions can only be obtained 
 again for asymptotic jet energies
when the kinematic boundaries can be ignored.
In that case, all but the path integral can be done
giving
\begin{eqnarray}
\Delta E 
 &\approx& \frac{C_R\alpha_s}{2}\int^\infty_{z_0} d z\,
\frac{\mu^2(z)}{\lambda(z)}\;(z-z_0) \log \frac{2E}{L\mu^2(z)}\; ,
\label{de11}\end{eqnarray} 
which is a linear weighed line integral over the local transport
coefficient\cite{bdms8} $(\mu^2(z)/\lambda(z))\approx
\frac{9}{2} \pi\alpha_s^2 \rho(z)$ however
enhanced by
a $\log 2E/L\mu^2(z)$ factor that results
from the structure of the GLV integral missing in the BDMS asymptotic limit. 
For an  expanding plasma as in~(\ref{rhoex})
\begin{eqnarray}
\Delta E_\alpha(L,z_0) 
&\approx&\frac{C_R\alpha_s}{2} \frac{\mu^2(L)L^\alpha}{\lambda(L)}\;
\frac{L^{2-\alpha}}{2-\alpha}\,\tilde{v} \;.
\label{deexp}\end{eqnarray}
Here $\tilde{v}=\log 2E/L\mu^2(L)$ and 
we used that $\mu^2(L)L^\alpha/\lambda(L)$ is
a constant independent of $L$ for this type of expansion.
We also took the $z_0\rightarrow 0$ limit. We therefore recover the
asymptotic BDMSZ energy loss for both static
 and  expanding media modulated by a $\log E/\omega(L)$ factor
that is important at RHIC energies.
Using the Bjorken relation between the
gluon density and the rapidity density then gives
\begin{eqnarray}
\Delta E_{\alpha=1}(L)&=& \
\frac{9C_R\pi\alpha_s^3}{4} \left(\frac{1}{\pi R^2} 
\frac{dN^g}{dy} \right) \, L\,\log \frac{2E}{L\mu^2(L)} \;. 
\label{debj}\end{eqnarray}
In practice, it is straight forward to
integrate GLV numerically including the 
finite kinematic constraints.

\begin{figure}
\begin{center}
\includegraphics[width=7cm,angle=270]{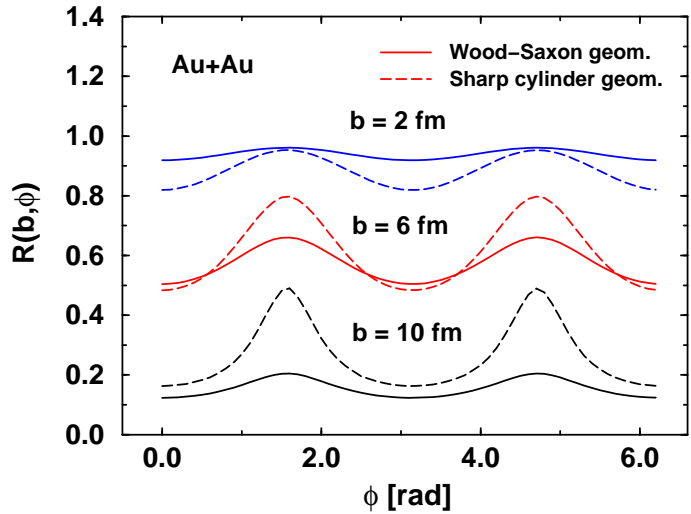}
     \caption{\small The modulation function  $R({\bf b},\phi)$   
is plotted vs. $\phi$ for impact parameters ${\bf b}=2, 6, 10$~fm.
Diffuse Wood-Saxon and uniform sharp cylinder geometries are compared. 
The most drastic difference between these geometries
occurs at  high impact parameters. }
\label{gvwfig3} 
\end{center}
\end{figure}

For non-central collisions the GLV line integral depends of course on the azimuthal direction $\phi$ of the jet. The variation of the azimuthal
energy loss with respect to $\phi$ at a given impact parameter $b$
can be expressed in terms of
$$R({\bf b},\phi )=\frac{\Delta E({\bf b},\phi )}{\Delta E(0)}$$
with results shown in Fig.\ref{gvwfig3} 
The effect of this azimuthal variation of the energy loss is to induce
an apparent elliptic flow at high $p_T$ not related to hydrodynamic
phenomena of low $p_T$. In \cite{gvw} we proposed a simple
interpolation between the hydrodynamic and jet quenched $p_T$
eikonal regimes
\begin{figure}
\begin{center}
\includegraphics[width=8cm,angle=270]{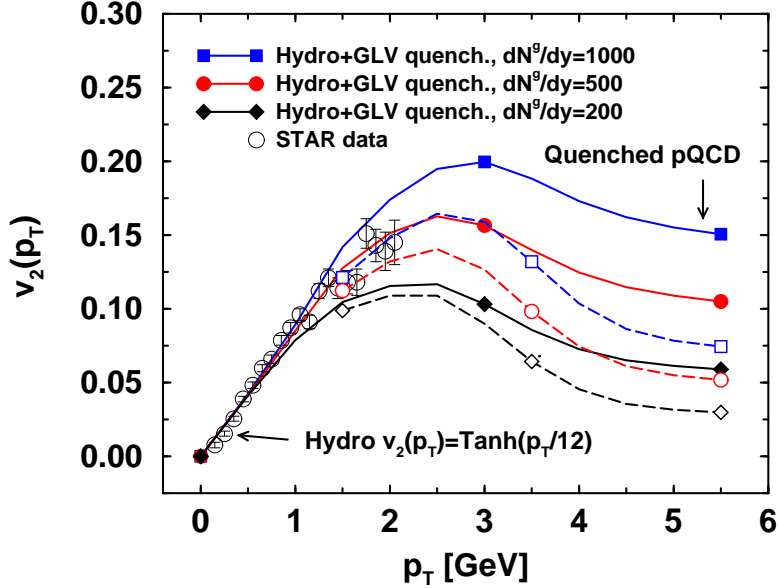}
       \caption{\small The interpolation of $v_2(p_{\rm T})$
between the soft hydrodynamic~\protect{\cite{huovinen}} and
hard pQCD regimes\protect\cite{gvw} 
is shown for different gluon rapidity densities in central
$b=0$ collisions. The gluon rapidity density at $b\ne 0$ is assumed
to scale with the binary collision number.
 Solid (dashed) curves
correspond to sharp cylindrical (diffuse Wood-Saxon) geometry}
\label{gvwfig4}
\end{center}
\end{figure}
\begin{equation}
v_{2}(p_{\rm T}) \approx 
\frac{v_{2s}(p_{\rm T}) dN_s + v_{2h}(p_{\rm T}) dN_h}
{ dN_s+dN_h } \;.
\label{v2interp}
\end{equation}
This interpolates between the hydrodynamic and the pQCD regimes because
at high $p_{\rm T}$, $dN_h\gg dN_s$. For our numerical estimates
the low $p_{\rm T}$ interpolation is achieved
by turning off the pQCD curves with a switch function 
$f_h({p}_{\rm T})= 0.5[1+{\rm tanh}(3 (p_{\rm T}-1.5\,{\rm GeV}))  ] $.

We see in Fig.\ref{gvwfig4}
that the magnitude and shape of the high 
$p_T$ elliptic flow provides a complementary
probe of the initial gluon density and  is
also sensitive to the geometrical
distribution of the plasma.
The saturated $v_2$ increases systematically
with increasing initial plasma density and thus  provides
an  important complementary
constraint on the maximum initial parton density produced
in central ${\bf b}=0$ collisions. The consistency of the quenched
elliptic flow in non-central with the central quench pattern
will be very important to test when the final data become available.

\section{Summary}

If confirmed by further measurements and theoretical
refinements, jet quenching may have already provided
the first evidence  
that initial parton densities 
on the order of 100 times nuclear matter density may have been
produced at RHIC.
The full analysis of flavor composition, shape, and azimuthal moments
of the high $p_T$ spectra appears to be 
a promising diagnostic probe of the
evolution of the gluon plasma produced at RHIC.
However, it is too early to tell what the preliminary
say about the properties of that extremely dense form of matter.
There are too many 
 pieces of the puzzle that simply do not fit well into 
any scenario. The beam energy and centrality independence 
of the transverse energy per charged particle is one of them.
The anomalous baryon number transport to high transverse momenta
and central rapidities is another.
Finally, the puzzling beam energy independence of the 
preliminary pion interferometry
results is  a mystery. As the tera-bytes of RHIC data continue
to stream
in during the next few years, they will certainly
pose many interesting new QCD
many body problems. The new chapter on the physics of ultra-dense matter
and the dynamics of ultra-relativistic nuclei is 
now unfolding at RHIC.

\section{Acknowledgments}

I especially thank T. Csorgo,
A. Dumitru, P. Levai, L. McLerran, D. Molnar, D. Rischke, and I. Vitev
for extensive collaborative work related to the topics of these lectures.
This work was supported by  
the Director, Office of Energy 
Research, Office of High Energy and Nuclear Physics, 
Division of Nuclear Physics, and by the Office of Basic Energy Science, 
Division of Nuclear Science, of  
the U.S. Department of Energy 
under Contract 
DE-FG02-93ER40764. I also thank the Collegium Budapest for hospitality
and partial support.

\end{document}